\documentclass[a4paper,11pt]{article}
\usepackage{graphicx,epsfig}
\usepackage{fancyhdr,fancybox}
\usepackage{indentfirst}
\usepackage{verbatim}
\usepackage[sort&compress, numbers]{natbib}
\usepackage{geometry}
\usepackage{extarrows,chemarrow,xypic} 
\usepackage[small]{caption2}
\usepackage{color}
\usepackage{microtype}
\DisableLigatures[f]{encoding = *, family = *}

\usepackage{times}     

\usepackage{amssymb}                 
\usepackage{amsthm}
\usepackage{bbm}
\usepackage{hyperref}  

\newcommand{\paperfont}{\fontsize{11pt}{1.2\baselineskip}\selectfont}
\begin{document}

\theoremstyle{definition}
\makeatletter
\thm@headfont{\bf}
\makeatother
\newtheorem{definition}{Definition}
\newtheorem{example}{Example}
\newtheorem{theorem}{Theorem}
\newtheorem{lemma}{Lemma}
\newtheorem{corollary}{Corollary}
\newtheorem{remark}{Remark}
\newtheorem{proposition}{Proposition}

\lhead{}
\rhead{}
\lfoot{}
\rfoot{}

\renewcommand{\refname}{References}
\renewcommand{\figurename}{Fig.}
\renewcommand{\tablename}{Table}
\renewcommand{\proofname}{Proof}

\newcommand{\diag}{\mathrm{diag}}
\newcommand{\one}{\mathbbm{1}}
\newcommand{\hyper}{{}_2F_1}
\newcommand{\confluent}{{}_1F_1}


\title{\textbf{Single-cell stochastic gene expression kinetics with coupled positive-plus-negative feedback}}
\author{Chen Jia$^{1,*}$, Le Yi Wang$^2$, George G. Yin$^1$, Michael Q. Zhang$^{3,4}$ \\
\footnotesize $^1$Department of Mathematics, Wayne State University, Detroit, MI 48202, U.S.A. \\
\footnotesize $^2$Department of Electrical and Computer Engineering, Wayne State University, Detroit, MI 48202, U.S.A. \\
\footnotesize $^3$Department of Biological Sciences, Center for Systems Biology, University of Texas at Dallas, Richardson, TX 75080, U.S.A. \\
\footnotesize $^4$MOE Key Laboratory of Bioinformatics, Center for Synthetic and Systems Biology, Tsinghua University, Beijing 100084, China. \\
\footnotesize $^*$Correspondence: chenjia@wayne.edu}
\date{}                              
\maketitle                           
\thispagestyle{empty}                

\paperfont

\begin{abstract}
Here we investigate single-cell stochastic gene expression kinetics in a minimal coupled gene circuit with positive-plus-negative feedback. A triphasic stochastic bifurcation upon the increasing ratio of the positive and negative feedback strengths is observed, which reveals a strong synergistic interaction between positive and negative feedback loops. We discover that coupled positive-plus-negative feedback amplifies gene expression mean but reduces gene expression noise over a wide range of feedback strengths when promoter switching is relatively slow, stabilizing gene expression around a relatively high level. In addition, we study two types of macroscopic limits of the discrete chemical master equation model: the Kurtz limit applies to proteins with large burst frequencies and the L\'{e}vy limit applies to proteins with large burst sizes. We derive the analytic steady-state distributions of the protein abundance in a coupled gene circuit for both the discrete model and its two macroscopic limits, generalizing the results obtained in [Chaos 26:043108, 2016]. We also obtain the analytic time-dependent protein distribution for the classical Friedman-Cai-Xie random bursting model proposed in [Phys. Rev. Lett. 97:168302, 2006]. Our analytic results are further applied to study the structure of gene expression noise in a coupled gene circuit and a complete decomposition of noise in terms of five different biophysical origins is provided.

\end{abstract}

\section{Introduction}
Gene expression in individual cells is an inherently stochastic process due to small copy numbers of biochemical molecules and probabilistic collisions between them \cite{kaern2005stochasticity}. Active genes are usually present in zero to one copy, mRNAs can be equally rare, and most proteins are present in less than 100 copies per bacterial cell \cite{paulsson2005models}. The simplest model of stochastic gene expression kinetics is the classical birth-death process \cite{feller1968introduction}, which characterizes the synthesis and degradation of mRNAs or proteins. The steady-state distribution for the birth-death process turns out to be a Poisson distribution, whose mean and variance are equal. However, numerous single-cell experiments have shown that the variance of expression levels is significantly larger than the mean for the majority of genes \cite{anders2010differential}, a phenomenon referred to as over-dispersion. To explain this observation, biologists managed to fit gene expression data with a negative binomial distribution \cite{robinson2010edger, anders2010differential} and biophysicists have justified the negative binomial distribution from the theoretical aspect \cite{paulsson2000random, shahrezaei2008analytical}.

Over the past two decades, numerous strides have been made in the single-cell biochemical reaction kinetics of stochastic gene expression \cite{peccoud1995markovian, paulsson2000random, kepler2001stochasticity, sasai2003stochastic, hornos2005self, friedman2006linking, raj2006stochastic, ramos2007symmetry, shahrezaei2008analytical, walczak2009stochastic, mugler2009spectral, iyer2009stochasticity, feng2010adiabatic, ramos2011exact, grima2012steady, radulescu2012relating, feng2012analytical, mackey2013dynamic, innocentini2013multimodality, pendar2013exact, jia2014modeling, kumar2014exact, yvinec2014adiabatic, kumar2015transcriptional, pajaro2015shaping, potoyan2015dichotomous, newby2015bistable, ge2015stochastic, liu2016decomposition, jkedrak2016time, popovic2016geometric, lin2016gene, lin2016bursting, veerman2017time, jia2017simplification, jia2017stochastic, jia2017emergent, bressloff2017stochastic, jia2018relaxation, lin2018efficient, jia2019analytic, chen2019limit}, which has a dual representation in terms of its probability distribution and stochastic trajectory. The former is usually described by a system of chemical master equations (CMEs) that is first introduced by Delbr\"{u}ck \cite{delbruck1940statistical}, while the latter is usually described by a continuous-time Markov jump process that can be computationally simulated via Gillespie's stochastic simulation algorithm. Readers may refer to \cite{schnoerr2017approximation} for recent comprehensive reviews about basic concepts and methods in this field.

The models of stochastic gene expression can be classified into two categories: discrete models and continuous models. The discrete models characterize the dynamics of the copy numbers of mRNAs and proteins. The first study of stochastic gene expression kinetics based on the discrete CME model was carried out by Berg \cite{berg1978model} and a thorough study was implemented by Shahrezaei and Swain \cite{shahrezaei2008analytical}. However, in bulk experiments and numerous single-cell experiments without single-molecule resolution such as single-cell RNA sequencing and flow cytometry, data are often recorded as continuous measurements at a macroscopic scale. These gene expression data boost the development of various continuous models, which characterize the dynamics of the concentrations (or densities) of mRNAs and proteins, copy numbers normalized by the system size.

Thus far, many continuous gene expression models have been proposed. Kepler et al. \cite{kepler2001stochasticity} modeled stochastic gene expression kinetics as a chemical Langevin equation. Friedman et al. \cite{friedman2006linking} proposed the continuous master equation model and it was pointed out later that the stochastic process underlying this model is a stochastic differential equation (SDE) driven by a compound Poisson process \cite{jkedrak2016time}. In addition, many authors \cite{newby2015bistable, ge2015stochastic, lin2016gene, bressloff2017stochastic, jia2017emergent} modeled stochastic gene expression kinetics as ordinary differential equations (ODEs) or SDEs with hybrid Markov switching in the regime of relatively slow promoter switching. In our recent work \cite{jia2017emergent}, we have unified most discrete and continuous models by regarding the latter as various macroscopic limits of the former. Under various discrete and continuous models, the steady-state and time-dependent probability distributions of mRNAs or proteins have been solved analytically by many authors (see \cite{veerman2017time} and the references therein). Recently, the CMEs for a wide class of gene regulatory networks have been solved approximately by Grima and coworkers under several linear approximations \cite{thomas2014phenotypic, cao2018linear}.

The previous gene expression models can also be classified according as the transcription dynamics is considered or not. Based on the central dogma of molecular biology, a complete model of stochastic gene expression should consider both transcription and translation. However, many previous papers focused more on the translation process and ignored the transcription process \cite{peccoud1995markovian, hornos2005self, grima2012steady, potoyan2015dichotomous, ge2015stochastic, jia2018relaxation}. In recent years, numerous single-cell experiments \cite{cai2006stochastic, suter2011mammalian} have shown that the synthesis of many mRNAs and proteins in individual cells may occur in random bursts | short periods of high expression intensity followed by long periods of low expression intensity | and it is known that random bursts of proteins result from short-lived mRNAs \cite{jia2017simplification}. Therefore, translational bursting cannot be fully described if the transcription dynamics is neglected.

The early work on stochastic gene expression focused on a simple transcription unit where the gene of interest is neither regulated by itself nor regulated by other genes \cite{paulsson2005models}. In recent years, many authors investigated an autoregulatory gene network with a positive or negative feedback loop \cite{hornos2005self}. In a recent work, Liu et al. \cite{liu2016decomposition} omitted the transcription step and studied stochastic gene expression kinetics in a minimal coupled gene circuit with both positive and negative feedback loops. Such kind of positive-plus-negative feedback networks widely exist in naturally occurring biological systems. In fact, they have been found in many bistable systems such as competence development in \emph{Bacillus subtilis} \cite{veening2008bistability} and many biological oscillators such as cell cycles and heartbeats \cite{tsai2008robust}.

In this paper, we present a detailed analysis of single-cell stochastic gene expression kinetics in a coupled gene circuit with promoter switching, transcription, translation, and positive-plus-negative feedback, extending the analytic results obtained in \cite{liu2016decomposition}. The structure of the present work is organized as follows. In Section 2, we introduce the three-stage CME model of a coupled gene circuit and study its two-time-scale model simplification. In Section 3, we derive the analytic steady-state distribution of the protein copy number in the presence or absence of translational bursting. In Section 4, we apply our analytic results to study the structure of gene expression noise in a coupled gene circuit and provide a complete decomposition of noise in terms of five different biophysical origins. Moreover, we observe a stochastic bifurcation upon the increasing ratio of the positive and negative feedback strengths, which reveals a crucial difference between coupled feedback loops and a single feedback loop. In particular, we discover that coupled positive-plus-negative feedback amplifies gene expression mean but reduces gene expression noise over a wide range of feedback strengths when promoter switching is relatively slow, stabilizing gene expression around a relatively high level. In Section 5, we investigate two types of macroscopic limits of the discrete CME model as the system size tends to infinity, which build a bridge between the discrete and continuous gene expression models proposed in previous papers. Finally, we derive the analytic steady-state distribution of the protein concentration under the two macroscopic limits and also obtain the analytic time-dependent protein distribution for the classical random bursting model proposed by Friedman et al. \cite{friedman2006linking}.

\section{Model}
\begin{figure}[!htb]
\centerline{\includegraphics[width=0.7\textwidth]{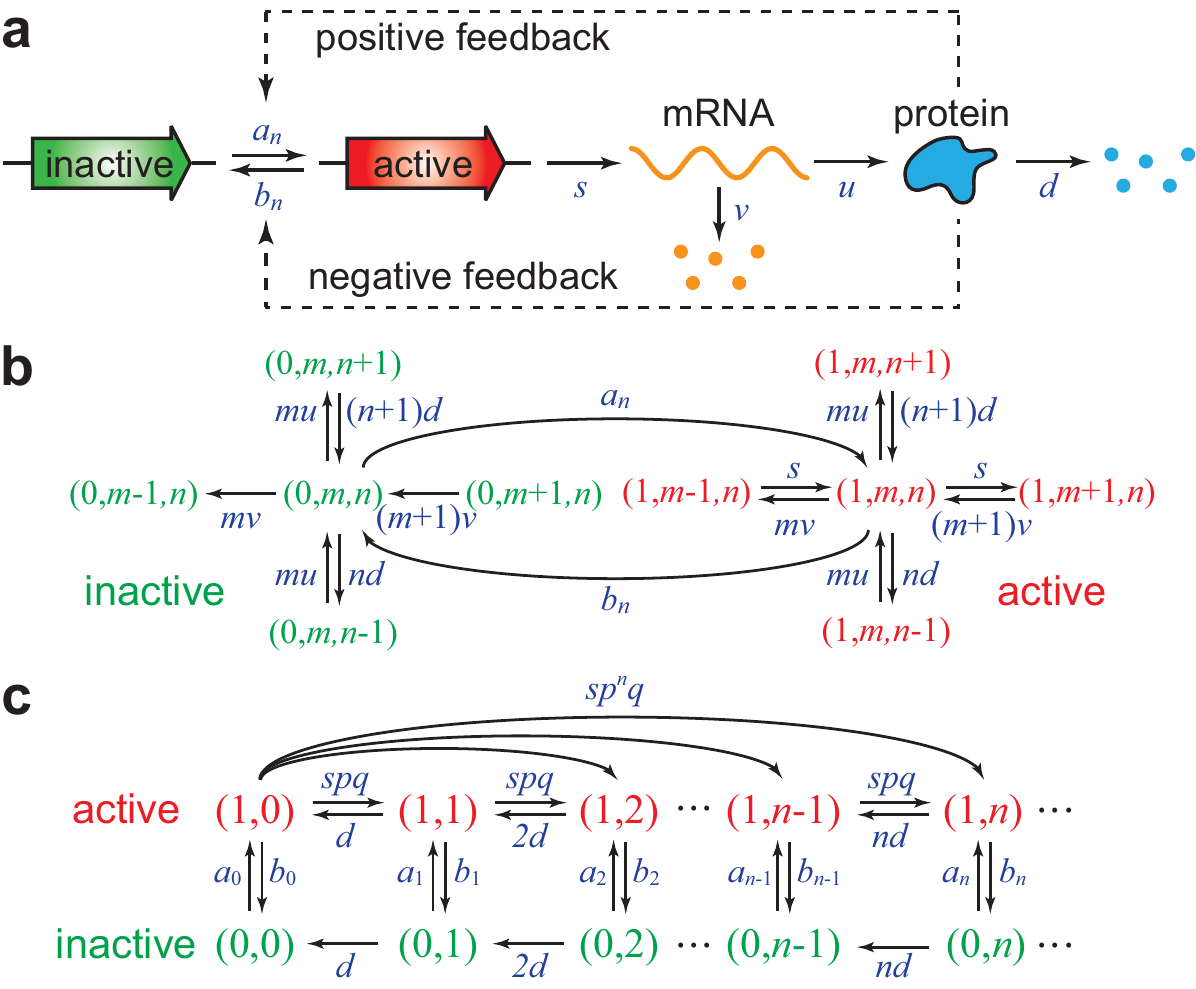}}
\caption{\textbf{A minimal coupled gene circuit with positive and negative feedback loops.} (a) The standard three-stage representation of stochastic gene expression consisting of transcription, and translation, and the switching of the promoter between an active and an inactive state. (b) The transition diagram of the full Markovian model whose evolution is governed by the CME. (c) The transition diagram of the reduced Markovian model when the mRNA decays much faster than the protein.}\label{model}
\end{figure}

Based on the central dogma of molecular biology, gene expression in a single cell has a standard three-stage representation involving transcription, translation, and switching of the promoter between an active and an inactive state (Fig. \ref{model}(a)) \cite{paulsson2005models, shahrezaei2008analytical}. The chemical reactions underlying the three-stage representation are listed as follows:
\begin{gather*}
\textrm{inactive gene} \xlongrightarrow{a_n} \textrm{active gene},\\
\textrm{active gene} \xlongrightarrow{b_n} \textrm{inactive gene},\\
\textrm{active gene} \xlongrightarrow{s} \textrm{active gene}+\textrm{mRNA},\;\;\;\\
\textrm{mRNA} \xlongrightarrow{u} \textrm{mRNA}+\textrm{protein},\\
\textrm{mRNA} \xlongrightarrow{v} \varnothing,\\
\textrm{protein} \xlongrightarrow{d} \varnothing.
\end{gather*}
where the first two reactions describe promoter switching, the middle two describe transcription and translation, and the last two describe the degradation of the mRNA and protein. The chemical state of the gene of interest can be represented by the ordered triple $(i,m,n)$: the activity $i$ of the promoter, the copy number $m$ of the mRNA, and the copy number $n$ of the protein. Here $i = 1$ and $i = 0$ correspond to the active and inactive states of the promoter, respectively.

Let $p_{i,m,n}(t)$ denote the probability of having $m$ copies of mRNA and $n$ copies of protein at time $t$ when the promoter is in state $i$. Then the dynamics of stochastic gene expression can be described by a continuous-time Markov jump process illustrated in Fig. \ref{model}(b). The evolution of the Markovian model is governed by the CME
\begin{equation*}\left\{
\begin{split}
\dot p_{0,m,n} =&\; (m+1)vp_{0,m+1,n}+mup_{0,m,n-1}\\
&\; +(n+1)dp_{0,m,n+1}+b_np_{1,m,n}\\
&\; -(mv+mu+nd+a_n)p_{0,m,n},\\
\dot p_{1,m,n} =&\; sp_{1,m-1,n}+(m+1)vp_{1,m+1,n}+mup_{1,m,n-1}\\
&\; +(n+1)dp_{1,m,n+1}+a_np_{0,m,n}\\
&\; -(s+mv+mu+nd+b_n)p_{1,m,n}.
\end{split}\right.
\end{equation*}
Here $s$ is the transcription rate; $u$ is the translation rate; $v$ and $d$ are the degradation rates of the mRNA and protein, respectively. In this paper, we consider a minimal coupled gene circuit with both positive and negative feedback loops, as illustrated in Fig. \ref{model}(a). Due to feedback regulation, the protein copy number $n$ will directly or indirectly affect the switching rates $a_n$ and $b_n$ of the promoter between the active and inactive states. Following \cite{kumar2014exact}, we assume that $a_n = a+\mu n$ and $b_n = b+\nu n$, where $a$ and $b$ are spontaneous switching rates of the promoter and $\mu$ and $\nu$ characterize the strengths of positive and negative feedback loops, respectively. In \cite{kumar2014exact}, the authors considered an autoregulatory gene circuit with either positive or negative feedback and thus one of $a_n$ and $b_n$ is a constant independent of $n$. Since the present work focuses on a coupled gene circuit, both $a_n$ and $b_n$ are functions of $n$.

Before further analysis, we stress that gene expression is a tremendously complex stochastic process that consists of many important chemical processes such as the binding of RNA polymerase (RNAP) to the promoter, the activation and repression of the RNAP-promoter complex by transcription factors, and transcriptional elongation. Each of these processes consists of a number of elementary reactions whose rates are strongly dependent on various cell-state variables including the concentrations of RNAP and transcription factors, the interaction strengths of genes with RNAP and transcription factors, the gene copy number, the phase of the cell cycle, the density of nutrients, and the microscopic conformation of chromosomes \cite{lim2015quantitative}. All of these cell-state variables are stochastic variables that differ from cell to cell and fluctuate over time. In particular, transcription at each gene state can be a non-Poisson, multi-step enzyme process whose rate is coupled to various cell-state variables. Although neglected in our model, the complex interactions of gene networks with hidden cell environments are important sources of gene expression noise in living cells \cite{park2018chemical}.

In single-cell experiments, it was consistently observed that mRNAs decay substantially faster than proteins \cite{shahrezaei2008analytical}. In fact, mRNA lifetimes in prokaryotes are usually on the order of a few minutes, while protein lifetimes are generally on the order of tens of minutes to many hours \cite{bernstein2002global}. This suggests that the process of protein synthesis followed by mRNA degradation is essentially instantaneous and thus protein synthesis will occur in random bursts. Since the mRNA kinetics is fast, the transcription process can be averaged out and the chemical state of the gene can be described by the ordered pair $(i,n)$. Once a transcript is synthesized, it can either produce a protein copy with probability $p = u/(u+v)$ or be degraded with probability $q = 1-p = v/(u+v)$. Therefore, the probability that a transcript can produce $k$ copies of protein before it is finally degraded is $p^kq$, which has a geometric distribution. The effective rate at which $k$ proteins are synthesized in a single burst will be the product of the transcription rate $s$ and the geometric probability $p^kq$. These considerations lead to the reduced Markovian model illustrated in Fig. \ref{model}(c) \cite{paulsson2000random, jia2017emergent}. In fact, the reduced model can be derived rigorously as the two-time-scale limit of the full model when $\lambda = v/d \gg 1$ and $u/v$ is finite. Readers interested in the rigorous mathematical theory may refer to \cite{jia2017simplification}.

\begin{figure}[!htb]
\centerline{\includegraphics[width=1\textwidth]{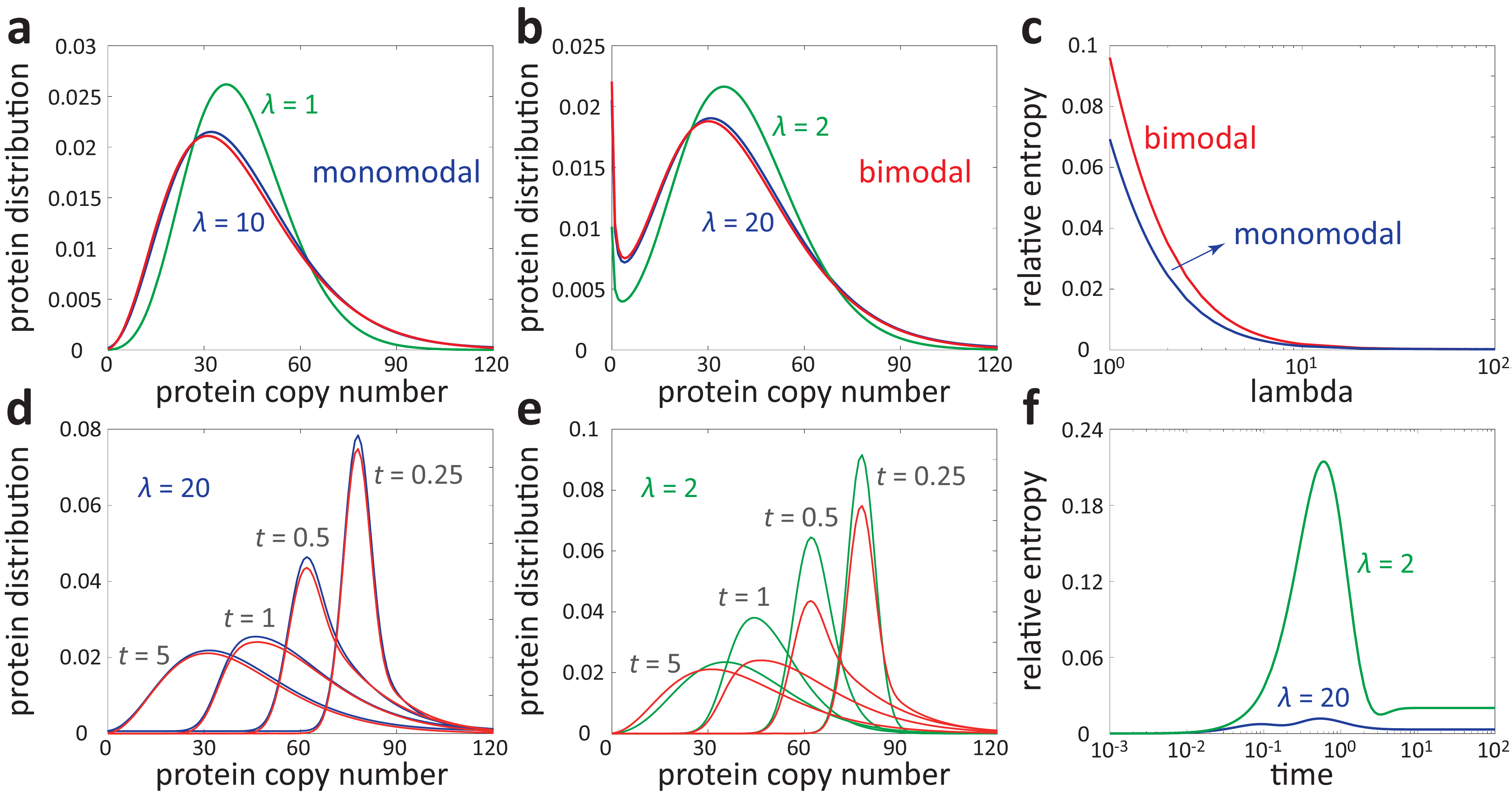}}
\caption{\textbf{Comparison of the full and reduced Markovian models.}
(a) Simulations of the steady-state protein distributions for the reduced model (red) and the full model when $\lambda = 1$ (green) and $\lambda = 10$ (blue). The protein distributions are monomodal.
(b) Simulations of the steady-state protein distributions for the reduced model (red) and the full model when $\lambda = 2$ (green) and $\lambda = 20$ (blue). The protein distributions are bimodal. In (a) and (b), the model parameters are chosen as $s = 5, d = 1, p = 0.9, q = 0.1$ and the promoter switching rates are chosen as $a_n = 5, b_n = 0.5$ in (a) and $a_n = 0.5+0.1n, b_n = 0.5$ in (b).
(c) Relative entropies between the steady-state protein distributions for the full and reduced models in the monomodal and bimodal cases.
(d) Simulations of the time-dependent protein distributions for the reduced (red) and full (blue) models under $\lambda = 20$ at different time points.
(e) Simulations of the time-dependent protein distributions for the reduced (red) and full (green) models under $\lambda = 2$ at different time points.
(f) Relative entropies between the time-dependent protein distributions for the full and reduced models versus the time under $\lambda = 2$ (green) and $\lambda = 20$ (blue). In (c)-(f), the model parameters are chosen as $s = 5, d = 1, p = 0.9, q = 0.1, a_n = 5, b_n = 0.5$.}\label{translation}
\end{figure}

To show the validity of the two-time-scale model simplification, we numerically simulate both the full and reduced Markovian models using Gillespie's algorithm. Fig. \ref{translation} (a),(b) illustrate the steady-state distributions of the protein copy number for the two models under two sets of biologically relevant parameters. Our model could yield monomodal or bimodal steady-state protein distributions. It has been shown in previous studies that bistability tends to occur in positive feedback networks \cite{jia2014modeling} and slow promoter switching could broaden the region of bistability \cite{ge2018relatively}. Therefore, the model parameters in the bimodal case are chosen in the regime of positive feedback and slow promoter switching. It can be seen that the steady-state protein distributions for the two models agree with each other perfectly when $\lambda\gg 1$, but they fail as expected for smaller $\lambda$. In the bimodal case, the reduced model may even reverse the heights of the two peaks when $\lambda$ is small (Fig. \ref{translation}(b)).

In statistical physics and probability theory, the relative entropy, also referred to as Kullback-Leibler divergence, is widely applied to characterize the similarity between two probability distributions. The relative entropy vanishes if and only if the two probability distributions are exactly the same. Fig. \ref{translation}(c) depicts the relative entropy between the steady-state protein distributions for the two models. It can be seen that the relative entropy decays dramatically when $\lambda$ is small and is close to zero when $\lambda\gg 1$. In addition, the model simplification performs better in the monomodal case and yields a larger error in the bimodal case, which needs a larger $\lambda$ to achieve the same approximation accuracy.

We next compare the dynamic properties of the full and reduced models. To this end, we illustrate the time-dependent distributions of the protein copy number for the two models in Fig. \ref{translation}(d),(e) when they start from the same initial distribution. When $\lambda\gg 1$, the two models exhibit almost the same dynamic behavior. However, the reduced model deviates from the full one in the small $\lambda$ regime. In addition, we depict the relative entropy between the time-dependent protein distributions for the two models in Fig. \ref{translation}(f), from which we can see that the protein distributions for the two models agree with each other reasonably well over the whole time axis when $\lambda\gg 1$.

Let $p_{i,n}(t)$ denote the probability of having $n$ copies of protein at time $t$ when the promoter is in state $i$. Then the evolution of the reduced Markovian model is governed by the CME
\begin{equation}\label{master}\left\{
\begin{split}
\dot p_{0,n} =&\; (n+1)dp_{0,n+1}+b_np_{1,n}-(nd+a_n)p_{0,n},\\
\dot p_{1,n} =&\; \sum_{k=0}^{n-1}sp^{n-k}qp_{1,k}+(n+1)dp_{1,n+1}\\
&\; +a_np_{0,n}-(sp+nd+b_n)p_{1,n}.
\end{split}\right.
\end{equation}
The reduced model contains long-range transitions of the protein copy number, which indicates that protein synthesis occurs in random bursts. The burst size of the protein is geometrically distributed and the mean burst size is given by $\sum_{k=0}^\infty kp^kq = p/q = u/v$.

\section{Steady-state protein distribution}

\subsection{Model with translational bursting}
An important question is whether the CME of a coupled gene circuit can be solved explicitly. In fact, the time-dependent solution has been obtained analytically for simple transcription units without feedback \cite{shahrezaei2008analytical, iyer2009stochasticity, feng2012analytical, pendar2013exact, jkedrak2016time} and has been obtained approximately for autoregulatory gene circuits when the feedback is weak \cite{iyer2014mixed, veerman2017time}. However, the time-dependent solution is difficult to obtain for a general coupled gene circuit with arbitrary feedback strengths. Here we study its steady-steady solution. To this end, we define a pair of generating functions
\begin{equation*}
f(z,t) = \sum_{n=0}^\infty p_{1,n}(t)z^n,\;\;\;g(z,t) = \sum_{n=0}^\infty p_{0,n}(t)z^n.
\end{equation*}
Moreover, let $p_n(t) = p_{1,n}(t)+p_{0,n}(t)$ denote the probability of having $n$ copies of protein at time $t$ and let $F(z,t) = f(z,t)+g(z,t)$ denote its generating function. At the steady state, both $f(z,t)$ and $g(z,t)$ are independent of time $t$ and the CME \eqref{master} can be transformed into the following system of ODEs:
\begin{equation*}\left\{
\begin{split}
& ag+[d(z-1)+\mu z]g'-bf-\nu zf' = 0,\\
& \left[\frac{sp(z-1)}{pz-1}+b\right]f+[d(z-1)+\nu z]f'-ag-\mu zg' = 0.
\end{split}\right.
\end{equation*}
The above system of ODEs has an explicit solution which is given by \cite[Section 1]{supp}
\begin{gather}\label{generating}
F(z) = \frac{\hyper(\alpha_1,\alpha_2;\beta;w(z-z_0))}{\hyper(\alpha_1,\alpha_2;\beta;w(1-z_0))},\\
f(z) = \frac{a(1-pz)}{\beta(\mu+\nu+dq)}
\frac{\hyper(\alpha_1+1,\alpha_2+1;\beta+1;w(z-z_0))}{\hyper(\alpha_1,\alpha_2;\beta;w(1-z_0))},\nonumber
\end{gather}
where $\hyper(\alpha_1,\alpha_2;\beta;z)$ is Gauss's hypergeometric function and
\begin{gather*}
\alpha_1+\alpha_2 = \frac{a+b}{\mu+\nu+d}+\frac{s(\mu+d)}{d(\mu+\nu+d)},\;\;\;
\alpha_1\alpha_2 = \frac{as}{d(\mu+\nu+d)},\\
\beta = \frac{a+b}{\mu+\nu+d}+\frac{sp\nu}{(\mu+\nu+d)(\mu+\nu+dq)},\;\;\;
w = \frac{p(\mu+\nu+d)}{\mu+\nu+dq},\;\;\;z_0 = \frac{d}{\mu+\nu+d}.
\end{gather*}
Then the steady-state distribution of the protein copy number can be recovered from $F$ as \cite[Equation 15.5.2]{special}
\begin{equation}\label{distribution}
p_n = \frac{F^{(n)}(0)}{n!} = \frac{(\alpha_1)_n(\alpha_2)_n}{(\beta)_n}\frac{w^n}{n!}
\frac{\hyper(\alpha_1+n,\alpha_2+n;\beta+n;-wz_0)}{\hyper(\alpha_1,\alpha_2;\beta;w(1-z_0))},
\end{equation}
where $(x)_n = x(x+1)\cdots(x+n-1)$ is the Pochhammer symbol. In addition, the steady-state mean of the protein copy number can be recovered from $F$ as \cite[Equation 15.5.1]{special}
\begin{equation}\label{mean}
\langle n\rangle = F'(1) = \frac{w\alpha_1\alpha_2}{\beta}
\frac{\hyper(\alpha_1+1,\alpha_2+1;\beta+1;w(1-z_0))}{\hyper(\alpha_1,\alpha_2;\beta;w(1-z_0))}.
\end{equation}
Although the full model has eight parameters $s,u,v,d,a,b,\mu,\nu$, the steady-state protein distribution only depends on five parameters $\alpha_1,\alpha_2,\beta,w,z_0$. This can be explained as follows. First, the averaging of the fast mRNA dynamics reduces a parameter. Next, it is clear that the steady-state distribution of a Markovian model remains the same if all transition rates are multiplied by a constant. This further reduces a parameter. Finally, since we focus on the steady-state solution rather than the time-dependent solution, this constraint also reduces a parameter.

Our analytic solution covers many results obtained in the previous literature. When $b = \mu = \nu = 0$, the switching from the active to the inactive state is forbidden and thus the gene is always active. In this case, the five parameters can be simplified as
\begin{equation*}
\alpha_1 = \beta = \frac{a}{d},\;\;\;\alpha_2 = \frac{s}{d},\;\;\;w = \frac{p}{q},\;\;\;z_0 = 1,
\end{equation*}
and thus the hypergeometric function in \eqref{distribution} reduces to \cite[Equation 15.4.6]{special}
\begin{equation*}
\hyper(\alpha_1+n,\alpha_2+n;\beta+n;-wz_0) = (1+wz_0)^{-(\alpha_2+n)} = q^{s/d+n}.
\end{equation*}
Therefore, the protein copy number has a negative binomial distribution
\begin{equation}
p_n = \frac{(s/d)_n}{n!}p^nq^{s/d},
\end{equation}
which is consistent with the result obtained by Paulsson and Ehrenberg \cite{paulsson2000random}. Moreover, the protein mean reduces to
\begin{equation}\label{simplemean}
\langle n \rangle_{\mathrm{active}} = \frac{s}{d}\times\frac{p}{q},
\end{equation}
where $s/d$ is the mean burst frequency \cite{friedman2006linking} and $p/q$ is the mean burst size. This quantity is often understood as the typical protein copy number in the active state \cite{assaf2011determining}. When $\mu = \nu = 0$, the promoter switching rates are constants and thus the gene is unregulated. In this case, our result coincides with the one obtained by Shahrezaei and Swain \cite{shahrezaei2008analytical}. When $\nu = 0$ or $\mu = 0$, the coupled gene circuit reduces to an autoregulatory gene circuit with positive or negative feedback and our result is in agreement with the one obtained by Kumar et al. \cite{kumar2014exact}.

When $a,b,\mu,\nu\gg s,d$, the promoter switches rapidly between the active and inactive states. In this case, the five parameters can be simplified as
\begin{equation*}
\alpha_1+\alpha_2 = \frac{a+b}{\mu+\nu}+\frac{s\mu}{d(\mu+\nu)},\;\;\;
\alpha_1\alpha_2 = \frac{as}{d(\mu+\nu)},\;\;\;
\beta = \frac{a+b}{\mu+\nu},\;\;\;w = p,\;\;\;z_0 = 0.
\end{equation*}
and thus the steady-state protein distribution reduces to
\begin{equation}\label{temp1}
p_n = A\frac{(\alpha_1)_n(\alpha_2)_n}{(\beta)_n}\frac{p^n}{n!},
\end{equation}
where $A = \hyper(\alpha_1,\alpha_2;\beta;w)^{-1}$ is a normalization constant. This is consistent with the result obtained by Mackey et al. \cite{mackey2013dynamic}. Since promoter switching is very fast, the gene states are in rapid pre-equilibrium and thus an effective transcription rate can be introduced as
\begin{equation}\label{effective}
c_n = \frac{sa_n}{a_n+b_n} = \frac{s(a+\mu n)}{(a+b)+(\mu+\nu)n},
\end{equation}
which has a Michaelis-Menten form. It is easy to check that
\begin{equation*}
\frac{(\alpha_1+n)(\alpha_2+n)}{\beta+n} = \frac{c_n}{d}+n.
\end{equation*}
Combining \eqref{temp1} and \eqref{effective}, the steady-state protein distribution can be rewritten as
\begin{equation}
p_n = A\frac{p^n}{n!}\frac{c_{0}}{d}\left(\frac{c_{1}}{d}+1\right)\cdots\left(\frac{c_{n-1}}{d}+n-1\right),
\end{equation}
which is consistent with the result obtained by Jia et al. \cite{jia2017stochastic}.

Another interesting question is to study the active probability of the gene. This quantity is important because it is closely related to the bursting dynamics of mRNAs \cite{jia2017simplification}. In fact, the steady-state probability that the gene is active can be recovered from $f$ as
\begin{equation}\label{activeprob}
P_{\mathrm{active}} =  f(1) = \frac{aq}{\beta(\mu+\nu+dq)}
\frac{\hyper(\alpha_1+1,\alpha_2+1;\beta+1;w(1-z_0))}{\hyper(\alpha_1,\alpha_2;\beta;w(1-z_0))}.
\end{equation}
When $\mu = \nu = 0$, the gene is unregulated. In this case, we have $z_0 = 1$ and $\beta = (a+b)/d$, and thus the active probability reduces to
\begin{equation*}
P_{\mathrm{active}} = \frac{a}{a+b}.
\end{equation*}
Interestingly, combining \eqref{mean} and \eqref{activeprob}, we obtain a universal relationship between the mean of the protein copy number and the active probability of the gene:
\begin{equation*}
\langle n\rangle = \frac{w\alpha_1\alpha_2(\mu+\nu+dq)}{aq}P_{\mathrm{active}}
= \frac{sp}{dq}P_{\mathrm{active}} = \langle n \rangle_{\mathrm{active}}P_{\mathrm{active}}.
\end{equation*}
This can be understood as follows. Recall that $\langle n \rangle_{\mathrm{active}}$ is the typical protein copy number in the active state, which can be understood as the conditional mean of the protein copy number given that the gene is active. This conditional mean, multiplied by the active probability of the gene, gives rise to the unconditional mean of the protein copy number.

\subsection{Model without translational bursting}
\begin{figure}[!htb]
\centerline{\includegraphics[width=0.6\textwidth]{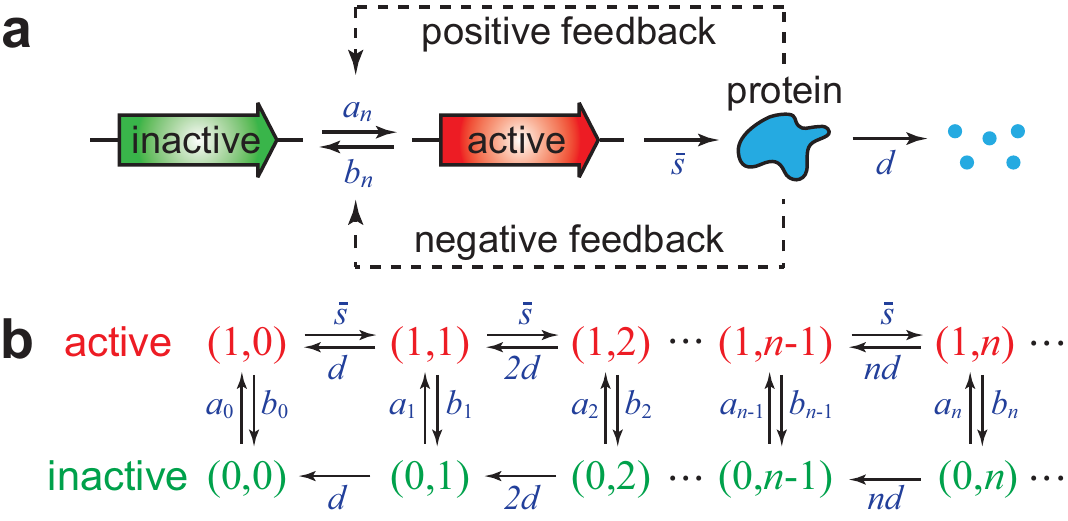}}
\caption{\textbf{A minimal coupled gene circuit without translational bursting.} (a) The two-stage representation of stochastic gene expression consisting of only promoter switching and translation, with the transcription step being ignored. (b) The transition diagram of the Markovian model describing the dynamics of the two-stage representation.}\label{nomRNA}
\end{figure}

There is another important case that should be paid special attention to. Consider the limiting case when $s\rightarrow\infty$ and $p\rightarrow 0$, while keeping $sp = \bar s$ as a constant. This is equivalent to assuming that the mean burst frequency $s/d\rightarrow\infty$ and the mean burst size $p/q\rightarrow 0$, while keeping their product $\langle n\rangle_{\mathrm{active}}$ as a constant. In this case, we have $q\rightarrow 1$ and thus
\begin{gather*}
spq \rightarrow \bar{s},\;\;\;sp^nq \rightarrow 0,\;\;\;n\geq 2.
\end{gather*}
Then the reduced model can be further simplified to a Markovian model without translational bursting, as depicted in Fig. \ref{nomRNA}(b). This model describes the dynamics of the two-stage representation of stochastic gene expression involving only promoter switching and translation, with the transcription process being ignored, as illustrated in Fig. \ref{nomRNA}(a).

To compute the generating function $F(z)$ in the absence of translational bursting, we recall that when $\alpha_1\rightarrow\infty$ and $z\rightarrow 0$, while keeping $\alpha_1z$ as a constant, Gauss's hypergeometric function has the following limit \cite[Equation 13.18.2]{special}:
\begin{equation*}
\hyper(\alpha_1,\alpha_2;\beta;z) \rightarrow \confluent(\alpha_2;\beta;\alpha_1z),
\end{equation*}
where $\confluent(\alpha;\beta;z)$ is Kummer's confluent hypergeometric function. Taking $s\rightarrow\infty$ and $p\rightarrow 0$ in \eqref{generating} and applying the above formula, the generating function $F(z)$ reduces to
\begin{equation*}
F(z) = \frac{\confluent(\alpha;\beta;w(z-z_0))}{\confluent(\alpha;\beta;w(1-z_0))},
\end{equation*}
where
\begin{equation*}
\alpha = \frac{a}{\mu+d},\;\;\;\beta = \frac{a+b}{\mu+\nu+d}+\frac{\bar s\nu}{(\mu+\nu+d)^2},\;\;\;
w = \frac{\bar s(\mu+d)}{d(\mu+\nu+d)},\;\;\;z_0 = \frac{d}{\mu+\nu+d}.
\end{equation*}
Then the steady-state protein distribution can be recovered from $F$ as \cite[Equation 13.3.15]{special}
\begin{equation}\label{confluent}
p_n = \frac{F^{(n)}(0)}{n!} = \frac{(\alpha)_n}{(\beta)_n}\frac{w^n}{n!}
\frac{\confluent(\alpha+n;\beta+n;-wz_0)}{\confluent(\alpha;\beta;w(1-z_0))}.
\end{equation}
This is consistent with the result obtained by Liu et al. \cite{liu2016decomposition}. Taking $s\rightarrow\infty$ and $p\rightarrow 0$ in \eqref{activeprob}, the steady-state active probability of the gene reduces to
\begin{equation}
P_{\mathrm{active}} = \frac{a}{\beta(\mu+\nu+d)}
\frac{\confluent(\alpha+1;\beta+1;w(1-z_0))}{\confluent(\alpha;\beta;w(1-z_0))}.
\end{equation}

The above analytic solution also covers some important results in previous studies. When the gene is always active, that is, $b = \mu = \nu = 0$, the four parameters can be simplified as
\begin{equation*}
\alpha = \beta = \frac{a}{d},\;\;\;w = \frac{\bar s}{d},\;\;\;z_0 = 1,
\end{equation*}
and thus the confluent hypergeometric function in \eqref{confluent} reduces to
\begin{equation*}
\confluent(\alpha+n;\beta+n;-wz_0) = e^{-wz_0} = e^{-\bar s/d}.
\end{equation*}
In this case, the protein copy number has a Poisson distribution:
\begin{equation}
p_n = \frac{(\bar s/d)^n}{n!}e^{-\bar s/d}.
\end{equation}
When $\mu = 0$ or $\nu = 0$, the coupled gene circuit reduces to an autoregulatory gene circuit with positive or negative feedback. In this case, our result is in agreement with the one obtained by Hornos et al. \cite{hornos2005self} and Grima et al. \cite{grima2012steady}. When the promoter switches rapidly between the active and inactive states, that is $a,b,\mu,\nu\gg\bar s,d$, the four parameters can be simplified as
\begin{gather*}
\alpha = \frac{a}{\mu},\;\;\;\beta = \frac{a+b}{\mu+\nu},\;\;\;
w = \frac{\bar s\mu}{d(\mu+\nu)},\;\;\;z_0 = 0.
\end{gather*}
and thus the steady-state protein distribution reduces to
\begin{equation}
p_n = A\frac{(\alpha)_n}{(\beta)_n}\frac{p^n}{n!},
\end{equation}
where $A = \confluent(\alpha;\beta;w)^{-1}$ is a normalization constant. This is consistent with the result obtained by Mackey et al. \cite{mackey2013dynamic}.

\section{Structure of gene expression noise and related stochastic bifurcation}
In the literature, the noise, also called dispersion, in the protein abundance within a cell population is often characterized by the Fano factor $\eta = \sigma^2/\langle n\rangle$, which is defined as the ratio of the variance $\sigma^2$ and the mean $\langle n\rangle$ \cite{friedman2006linking}. A dispersion greater than one reveals a deviation from the Poisson distribution and thus serves as a characteristic signal of over-dispersion. In fact, both the steady-state protein mean and protein noise can be recovered from $F$ as
\begin{equation}
\langle n\rangle = F'(1),\;\;\;\eta = \frac{F''(1)+F'(1)-F'(1)^2}{F'(1)}.
\end{equation}
Applying these formulas gives rise to
\begin{gather}
\langle n\rangle = \frac{w\alpha_1\alpha_2}{\beta} \frac{\hyper(\alpha_1+1,\alpha_2+1;\beta+1;w(1-z_0))}{\hyper(\alpha_1,\alpha_2;\beta;w(1-z_0))},\\
\eta = \frac{w(\alpha_1+1)(\alpha_2+1)}{\beta+1} \frac{\hyper(\alpha_1+2,\alpha_2+2;\beta+2;w(1-z_0)}{\hyper(\alpha_1+1,\alpha_2+1;\beta;w(1-z_0))}
+1-\langle n\rangle.\nonumber
\end{gather}
Based on a detailed analysis of hypergeometric functions, it can be shown that when $\mu,\nu\ll d$, the steady-state protein mean has the following approximation \cite[Section 2]{supp}:
\begin{equation}\label{decompositionmean}
\langle n\rangle \approx \frac{a}{a+b}\langle n\rangle_{\mathrm{active}}+
\langle n\rangle_{\mathrm{positive}}-\langle n\rangle_{\mathrm{negative}},
\end{equation}
where
\begin{equation*}
\langle n\rangle_{\mathrm{positive}} = \left[\frac{ab\langle n\rangle_{\mathrm{active}}^2}{(a+b)^2(a+b+d)}\right]\mu,\;\;\;
\langle n\rangle_{\mathrm{negative}} = -\left[\frac{a(a+d)\langle n\rangle_{\mathrm{active}}^2}{(a+b)^2(a+b+d)}\right]\nu.
\end{equation*}
Here $\langle n\rangle_{\mathrm{positive}}$ and $-\langle n\rangle_{\mathrm{negative}}$ are the contributions of positive and negative feedback loops to the protein mean, respectively. The former describes the effect of mean amplification caused by positive feedback, while the latter describes the effect of mean reduction caused by negative feedback. In other words, in a coupled gene circuit, positive feedback amplifies the protein mean, while negative feedback reduces it.

Similarly, when $\mu,\nu\ll d$, the steady-state protein noise can be decomposed into five different terms as \cite[Section 2]{supp}
\begin{equation}
\eta \approx \eta_{\mathrm{protein}}+\eta_{\mathrm{mRNA}}+\eta_{\mathrm{gene}}+
\eta_{\mathrm{positive}}-\eta_{\mathrm{negative}},\\
\end{equation}
where
\begin{gather*}
\eta_{\mathrm{protein}} = 1,\;\;\;
\eta_{\mathrm{mRNA}} = \frac{p}{q},\;\;\;
\eta_{\mathrm{gene}} = \frac{a+d}{a+b+d}\langle n\rangle_{\mathrm{active}}-\langle n\rangle,\\
\eta_{\mathrm{positive}} = \frac{\langle n\rangle_{\mathrm{active}}}{a+b+d} \left\{\frac{b}{a+b+d}+\frac{bp}{(a+b+2d)q}\left[1+\frac{s(a+d)}{d(a+b+d)}\right]\right\}\mu,\\
\eta_{\mathrm{negative}} = \frac{\langle n\rangle_{\mathrm{active}}}{a+b+d} \left\{\frac{a+d}{a+b+d}+\frac{(a+2d)p}{(a+b+2d)q}\left[1+\frac{s(a+d)}{d(a+b+d)}\right]\right\}\nu.
\end{gather*}
Here $\eta_{\mathrm{protein}} = 1$ is the Poisson noise from individual births and deaths of the protein, $\eta_{\mathrm{mRNA}} = p/q$, which is equal to the mean burst size of the protein, characterizes fluctuations in the mRNA abundance, and $\eta_{\mathrm{gene}}$ characterizes fluctuations due to promoter switching. Moreover, $\eta_{\mathrm{positive}}$ and $-\eta_{\mathrm{negative}}$ are the contributions of positive and negative feedback loops to the protein noise, respectively. The former describes the effect of noise amplification caused by positive feedback, while the latter describes the effect of noise suppression caused by negative feedback. Compared with similar decompositions in previous studies \cite{shahrezaei2008analytical}, our decomposition reveals five different biophysical origins of the protein noise and is very general. It clearly explains previous experimental observations that positive feedback amplifies gene expression noise \cite{becskei2001positive}, while negative feedback reduces it \cite{becskei2000engineering}. Moreover, it provides novel insights into how and to what extent coupled feedback loops can enhance or suppress molecular fluctuations.

We emphasize that our decompositions of the protein mean and protein noise are only valid in the regime of $\mu,\nu\ll d$, which has also been assumed in \cite{veerman2017time}. In fact, this assumption is satisfied over a wide range of biological systems. To see this, we notice that the feedback contribution $\mu n$ to the promoter switching rate usually has the same order as the spontaneous contribution $a$. This suggests that $a/\mu$ should have the same order as $\langle n\rangle_{\mathrm{active}} = sp/dq$ and thus $d/\mu$ and $sp/aq$ should have the same order. Similarly, $d/\nu$ and $sp/bq$ should have the same order. In living cells, the mean burst size of the protein, $p/q$, is relatively large, typically on the order of 100 for an \emph{E. coli} gene \cite{paulsson2005models}. In addition, recent single-cell experiments on transcription bursts of mammalian cells have shown that both $s/b$ and $s/a$ are also relatively large. In \cite{suter2011mammalian}, the authors monitored the transcription dynamics in mouse fibroblasts using single-cell time-lapse bioluminescence imaging and found that the three parameters $a$, $b$, and $s$ for different genes are typically on the order of 0.01/min, 0.1/min, and 1/min, respectively (see Figs. 1(D), 1(E), and S8 of \cite{suter2011mammalian} for details). These experimental measurements imply that both $d/\mu$ and $d/\nu$ are usually very large in real biological systems, which coincides with our assumption of $\mu,\nu\ll d$.

We next focus on two special cases. When $\mu = \nu = 0$, the gene is unregulated. In this case, the protein mean reduces to
\begin{equation*}
\langle n\rangle = \frac{w\alpha_1\alpha_2}{\beta} = \frac{a}{a+b}\langle n\rangle_{\mathrm{active}}
\end{equation*}
and the protein noise can be decomposed into three different terms as
\begin{equation}
\eta = \eta_{\mathrm{protein}}+\eta_{\mathrm{mRNA}}+\eta_{\mathrm{gene}},
\end{equation}
where the promoter switching noise $\eta_{\mathrm{gene}}$ can be computed explicitly as
\begin{equation*}
\eta_{\mathrm{gene}} = \frac{bd}{a(a+b+d)}\langle n\rangle.
\end{equation*}
This is fully consistent with the decomposition obtained by Shahrezaei and Swain \cite{shahrezaei2008analytical}.

When $a,b\gg s,d$, the promoter switches rapidly between the active and inactive states. In this case, the promoter switching noise $\eta_{\mathrm{gene}}$ is averaged out and the protein noise can be decomposed into four different terms as
\begin{equation}
\eta = \eta_{\mathrm{protein}}+\eta_{\mathrm{mRNA}}+\eta_{\mathrm{positive}}-\eta_{\mathrm{negative}}.
\end{equation}
This decomposition is consistent with the one obtained by Jia et al. \cite{jia2017stochastic} in the regime of fast promoter switching:
\begin{equation}
\eta = \eta_{\mathrm{protein}}+\eta_{\mathrm{mRNA}}+\frac{\mathrm{Cov}(n,c_n)}{\langle c_n\rangle},
\end{equation}
where $c_n$ is the effective transcription rate defined in \eqref{effective} and the third term is the relative covariance between $n$ and $c_n$. In the positive feedback case, $c_n$ is an increasing function of $n$ and the covariance term must be positive. In the negative feedback case, $c_n$ is an decreasing function of $n$ and the covariance term must be negative.

We next study the behavior of a coupled gene circuit as the feedback strengths $\mu$ and $\nu$ vary. It is easy to show that
\begin{equation}
\frac{\langle n\rangle_{\mathrm{positive}}}{\langle n\rangle_{\mathrm{negative}}} = \frac{\mu}{\delta_1\nu},\;\;\;
\frac{\eta_{\mathrm{positive}}}{\eta_{\mathrm{negative}}} = \frac{\mu}{\delta_2\nu},
\end{equation}
where $\delta_1$ and $\delta_2$ are two critical values given by
\begin{equation*}
\delta_1 = \frac{a+d}{b} < \delta_2 = \frac{\frac{a+d}{a+b+d}+\frac{(a+2d)p}{(a+b+2d)q}\left[1+\frac{s(a+d)}{d(a+b+d)}\right]}
{\frac{b}{a+b+d}+\frac{bp}{(a+b+2d)q}\left[1+\frac{s(a+d)}{d(a+b+d)}\right]}.
\end{equation*}
It is clear that $\delta_1<\delta_2<2\delta_1$. The two critical values $\delta_1$ and $\delta_2$ separate the parameter region into three phases, leading to a stochastic bifurcation. When $\mu/\nu<\delta_1$, both the protein mean and protein noise are reduced and the coupled gene circuit behaves as a negative feedback circuit. When $\mu/\nu>\delta_2$, both the protein mean and protein noise are amplified and the coupled gene circuit behaves as a positive feedback circuit. In the transitional phase of $\delta_1<\mu/\nu<\delta_2$, however, the protein mean is amplified but the protein noise is reduced. In this case, the coupled gene circuit behaves neither as a positive feedback nor as a negative feedback circuit and thus the positive and negative feedback effects cannot be cancelled out. The existence of the transitional phase reveals a crucial difference between coupled positive-plus-negative feedback loops and a single feedback loop.

\begin{figure}[!htb]
\centerline{\includegraphics[width=1\textwidth]{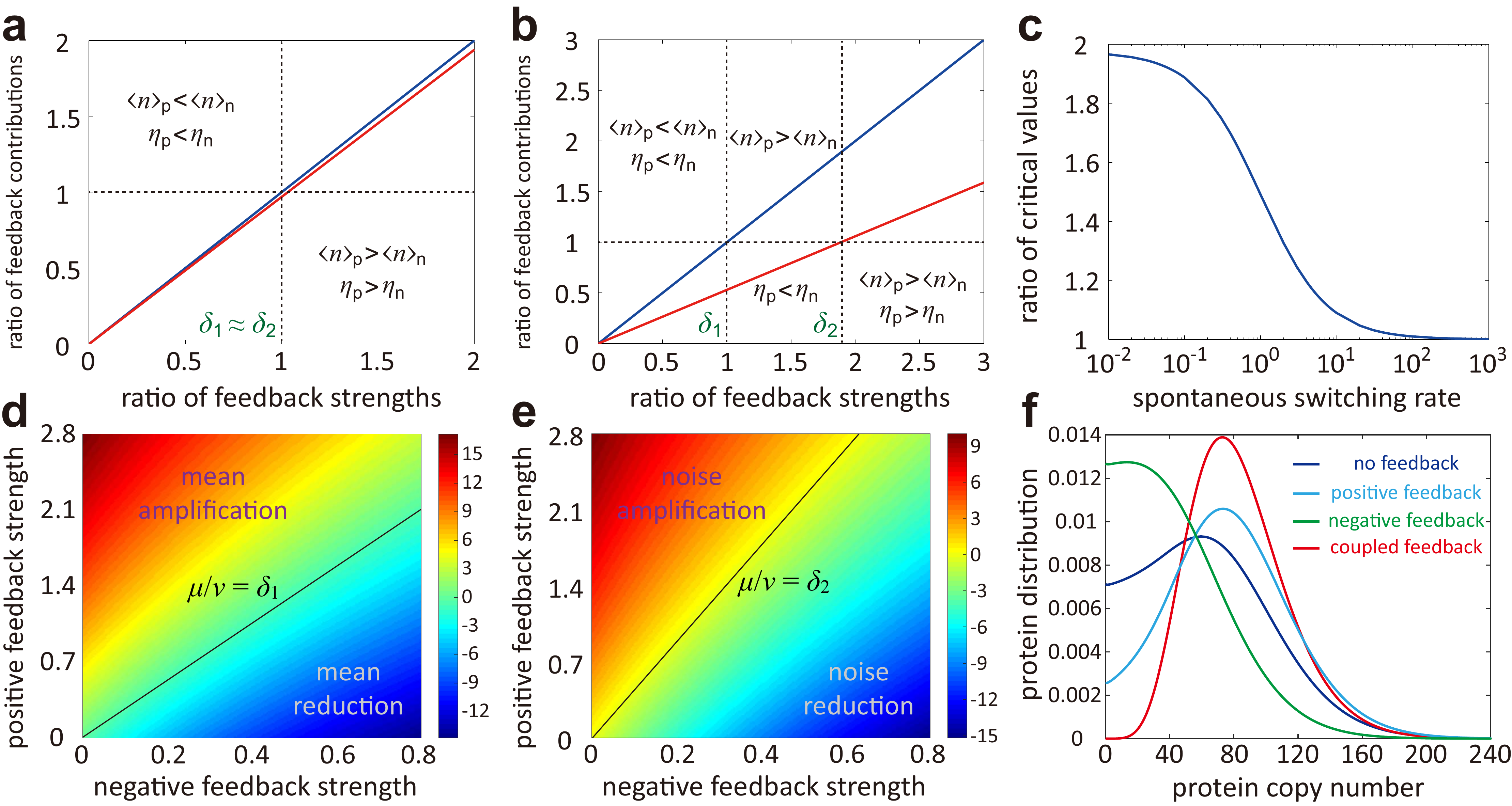}}
\caption{\textbf{Stochastic bifurcation in coupled gene circuits.} (a)-(b) The ratios of the positive and negative feedback contributions to the protein mean and protein noise, $\langle n\rangle_{\mathrm{positive}}/\langle n\rangle_{\mathrm{negative}}$ (blue) and $\eta_{\mathrm{positive}}/\eta_{\mathrm{negative}}$ (red), as functions of $\mu/\nu$. (a) The regime of fast promoter switching. (b) The regime of slow promoter switching. (c) The ratio of the two critical values $\delta_2/\delta_1$ versus the spontaneous switching rate $a$. The model parameters are chosen as $s = 10, d = 1, b = a+d, p/q = 10$ in (a)-(c), $a = 30$ in (a), and $a = 0.1$ in (b). (d) The heat map of the total feedback contribution to the protein mean, $\langle n\rangle_{\mathrm{positive}}-\langle n\rangle_{\mathrm{negative}}$, versus $\mu$ and $\nu$. (e) The heat map of the total feedback contribution to the protein noise, $\eta_{\mathrm{positive}}-\eta_{\mathrm{negative}}$, versus $\mu$ and $\nu$. (f) Steady-state protein distributions in four types of gene networks: simple transcript units without feedback (blue), positive feedback circuits (cyan), negative feedback circuits (green), and coupled gene circuits (red). The model parameters are chosen as $s = 10, d = 1, a = 1.06, b = 0.6, p/q = 10$ in (d)-(f). In (f), the feedback strengths $\mu$ and $\nu$ are chosen as $\mu = 0.02, \nu = 0$ in the positive feedback case, $\mu = 0, \nu = 0.015$ in the negative feedback case, and $\mu = 5, \nu = 1$ in the coupled feedback case.}\label{interaction}
\end{figure}

A special case occurs when promoter switching is very fast. In this case, we have $a,b\gg d$ and thus
\begin{equation*}
\delta_1 \approx \delta_2 \approx \frac{a}{b}.
\end{equation*}
Since the two critical values are very close, the transitional phase is almost invisible. When $\mu/\nu<a/b$, the coupled gene circuit behaves as a negative feedback circuit. When $\mu/\nu>a/b$, the coupled gene circuit behaves as a positive feedback circuit. To gain an intuitive picture of the stochastic bifurcation, we depict $\langle n\rangle_{\mathrm{positive}}/\langle n\rangle_{\mathrm{negative}}$ and $\eta_{\mathrm{positive}}/\eta_{\mathrm{negative}}$ as functions of $\mu/\nu$ in Fig. \ref{interaction}(a),(b). In the regime of fast promoter switching, the transitional phase is almost invisible and the coupled gene circuit behaves either as a positive feedback or as a negative feedback circuit. In the regime of slow promoter switching, we have $a,b\ll d$ and $\delta_2\approx 2\delta_1$. In this case, the transitional phase becomes much wider, which reveals a strong synergistic interaction between positive and negative feedback loops over a wide range of feedback strengths. Fig. \ref{interaction}(c) depicts the ratio of the two critical values $\delta_2/\delta_1$ versus the spontaneous switching rate $a$. It is easy to see that $\delta_2/\delta_1\rightarrow 1$ in the limit of $a\rightarrow\infty$, corresponding to fast promoter switching, while $\delta_2/\delta_1\rightarrow 2$ in the limit of $a\rightarrow 0$, corresponding to slow promoter switching. This again shows that the transitional phase becomes much wider as promoter switching becomes slower.

Our model predicts that coupled positive-plus-negative feedback amplifies gene expression mean but diminishes gene expression noise over a wide range of feedback strengths when promoter switching is relatively slow. This is clearly seen from Fig. \ref{interaction}(d),(e), which depict the heat maps of the total feedback contributions to the protein mean and protein noise, $\langle n\rangle_{\mathrm{positive}}-\langle n\rangle_{\mathrm{negative}}$ and $\eta_{\mathrm{positive}}-\eta_{\mathrm{negative}}$, versus the feedback strengths $\mu$ and $\nu$. Compared with a negative feedback circuit which stabilizes gene expression around a relatively low level and a positive feedback circuit which does not stabilize gene expression, a coupled gene circuit could stabilize gene expression around a relatively high level, as illustrated in Fig. \ref{interaction}(f).

Thus far, our predictions are made under the assumption that the feedback strengths are small, that is, $\mu,\nu\ll d$. However, according to our numerical simulations, our main results are actually insensitive to the feedback strengths. In particular, the stochastic bifurcation is also observed when the feedback strengths are relatively large (Fig. \ref{interaction}(d),(e)). We anticipate that our predictions could be validated in the near future via single-cell gene expression data.

\section{Macroscopic limits of stochastic gene expression kinetics}

\subsection{Kurtz limit}
In many single-cell experiments such as flow cytometry and fluorescence microscopy, one usually obtains data of protein concentrations, instead of protein copy numbers. Let $x = n/K$ be a continuous variable representing the protein concentration (or density), where $K$ is a large parameter with $K\rightarrow\infty$ corresponding to a macroscopic scale. In some previous papers, the parameter $K$ is chosen to be the average cell volume \cite{cai2006stochastic, taniguchi2010quantifying}. In the present paper, however, we follow the idea in \cite{assaf2011determining, lv2014constructing, ge2015stochastic} and choose $K\propto\langle n\rangle_{\mathrm{active}}$ to be an arbitrary quantity that is proportional to the typical protein copy number in the active state, which is usually very large in living cells. As $K\rightarrow\infty$, the concentration variable $x$ becomes continuous and the discrete stochastic gene expression kinetics has a macroscopic limit. Since
\begin{equation*}
\langle n\rangle_{\mathrm{active}} = \frac{s}{d}\times\frac{p}{q},
\end{equation*}
there are two different scenarios: If the mean burst frequency $s/d\rightarrow\infty$ while keeping the mean burst size $p/q$ as a constant, the resulting limit is called the Kurtz limit; If the mean burst size $p/q\rightarrow\infty$ while keeping the mean burst frequency $s/d$ as a constant, the resulting limit is called the L\'{e}vy limit \cite{jia2017emergent}.

\begin{figure}[!htb]
\centerline{\includegraphics[width=1\textwidth]{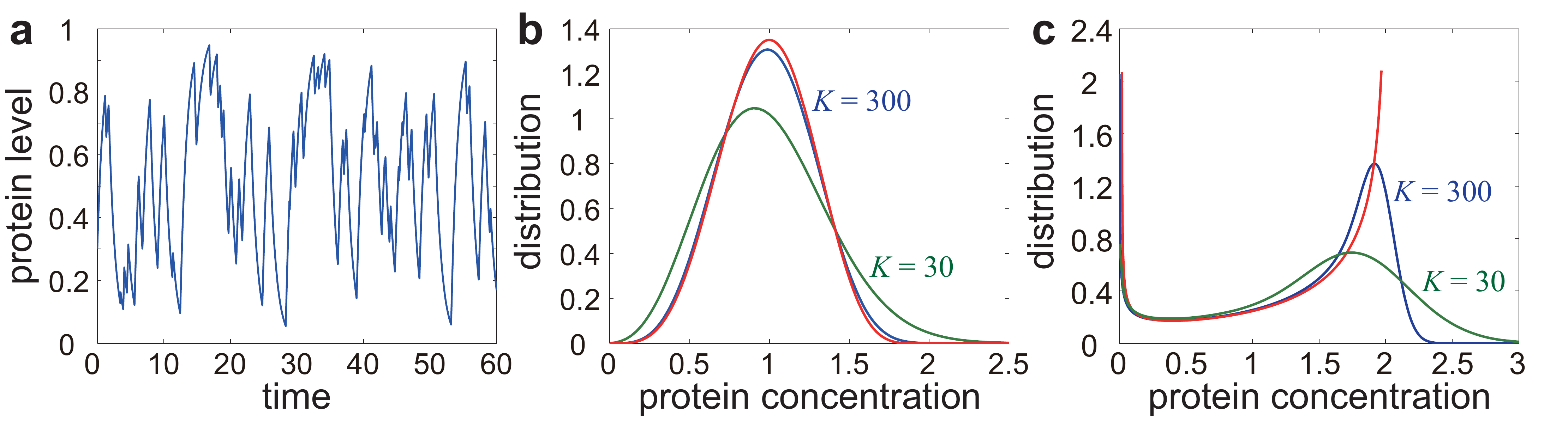}}
\caption{\textbf{Stochastic gene expression kinetics described by the Kurtz limit.}
(a) The simulated trajectory of the switching ODE model. Given a particular promoter state, the system evolves as an ODE with no fluctuations. The model parameters are chosen as $s' = 1, d = 1, p = 0.5, a = b = 1, \mu' = \nu' = 0.5$.
(b)-(c) Simulations of the steady-state distributions of the protein concentration for the switching ODE model (red) and the discrete CME model when $K = 30$ (green) and $K = 300$ (blue). (b) The case of monomodal protein distributions with the model parameters being chosen as $s' = 2, d = 1, p = 0.5, a = b = 5, \mu' = \nu' = 1$. (c) The case of bimodal protein distributions with the model parameters being chosen as $s' = 2, d = 1, p = 0.5, a = b = 0.5, \mu' = 1, \nu' = 0$.}\label{kurtz}
\end{figure}

We first investigate the Kurtz limit of the discrete CME model. To this end, we assume that the transcription rate scales with $K$ as $s = s'K$ and the feedback strengths scale with $1/K$ as $\mu = \mu'/K$ and $\nu = \nu'/K$, where we treat $s',d,p,a,b,\mu',\nu'$ as constants. Let $p_i(x,t)$ denote the probability density of the protein concentration at time $t$ when the promoter is in state $i$. When $K\gg 1$, the probability density $p_i(x,t)$ of the protein concentration and the probability distribution $p_{i,n}(t)$ of the protein copy number are related by
\begin{equation}\label{relation}
p_i(n/K,t) \approx Kp_{i,n}(t).
\end{equation}
Applying this relation and taking the limit of $K\rightarrow\infty$ in the CME \eqref{master}, we obtain the following system of partial differential equations \cite[Section 3]{supp}:
\begin{equation}\label{Kolmogorov}\left\{
\begin{split}
\partial_tp_0(x) &= d\partial_x(xp_0(x))+(b+\nu'x)p_1(x)-(a+\mu'x)p_0(x),\\
\partial_tp_1(x) &= d\partial_x(xp_1(x))-(s'p/q)\partial_xp_1(x)+(a+\mu'x)p_0(x)-(b+\nu'x)p_1(x).
\end{split}\right.
\end{equation}
From the viewpoint of stochastic processes, this is the Kolmogorov forward equation of the following switching ODE model:
\begin{equation*}
\xymatrix{
\dot{x} = s'p/q-dx \ar@<0.2ex>@^{->}[d]^{b+\nu'x} & (\textrm{active gene}),\\
\dot{x} = -dx \ar@<0.2ex>@^{->}[u]^{a+\mu'x} & (\textrm{inactive gene}).}
\end{equation*}
Therefore, the Kurtz limit of the discrete CME model is a switching ODE model, which is a special case of the so-called piecewise deterministic Markov process \cite{davis1984piecewise}. This is called the Kurtz limit because it is consistent with the classical Kurtz's limit theory of mesoscopic chemical reaction kinetics \cite{kurtz1972relationship}: given a particular gene state, the protein concentration evolves as an ODE with no fluctuations and thus all stochasticity comes from promoter switching. Fig. \ref{kurtz}(a) illustrates the simulated time series of the protein concentration in the Kurtz limit under a set of biologically relevant parameters, from which we can see that the stochastic trajectories of the switching ODE model are continuous. The increasing parts in the trajectory correspond to protein synthesis, while the decreasing parts correspond to protein degradation. When $\nu = 0$, the coupled gene circuit reduces to a circuit with positive autoregulation. In this case, Lin and Doering \cite{lin2016gene} also obtained a switching ODE by assuming that there is at most one copy of mRNA in a single cell with $m = 1$ corresponding to the active state and $m = 0$ corresponding to the inactive state. Compared to that work, our derivation is mathematically more rigorous.

\subsection{Steady-state protein distribution in the Kurtz limit}
We next study the steady-state protein distribution for the switching ODE model of a coupled gene circuit. Let $p(x) = p_1(x)+p_0(x)$ denote the steady-state probability density of the protein concentration and let $\hat{p}(\lambda) = \int_0^\infty p(x)e^{-\lambda x}dx$ denote its Laplace transform. We make a crucial observation that the generating function $F(z)$ will converge to the Laplace transform $\hat{p}(\lambda)$ as $z\rightarrow 1$ and $K\rightarrow\infty$ while keeping $\lambda = (1-z)K$ as a constant:
\begin{equation*}
F(z) = \sum_{n=0}^\infty p_ne^{n\log z} \approx \sum_{n=0}^\infty p_ne^{-n(1-z)}
\approx \sum_{n=0}^\infty p\left(\tfrac{n}{K}\right)e^{-\lambda\tfrac{n}{K}}\tfrac{1}{K} \rightarrow \int_0^\infty p(x)e^{-\lambda x}dx =  \hat{p}(\lambda),
\end{equation*}
where we have used the relation \eqref{relation} and the fact that a Riemann sum converges to a Riemann integral as the partition size tends to zero. Recall that Gauss's hypergeometric function and Kummer's confluent hypergeometric function are related by \cite[Equation 13.18.2]{special}
\begin{equation*}
\lim_{K\rightarrow\infty}\hyper(\alpha_1K,\alpha_2;\beta;z/K) = \confluent(\alpha_2;\beta;\alpha_1z).
\end{equation*}
Taking $z\rightarrow 1$ and $K\rightarrow\infty$ in the generating function \eqref{generating} and applying the above formula, we obtain the Laplace transform of the steady-state protein distribution \cite[Section 3]{supp}
\begin{equation}
\hat{p}(\lambda) = \frac{\confluent(\alpha;\beta;-w(\lambda-\lambda_0))}{\confluent(\alpha;\beta;w\lambda_0)},
\end{equation}
where
\begin{equation*}
\alpha = \frac{a}{d},\;\;\;\beta = \frac{a+b}{d}+\frac{s'\nu'p}{d^2q},\;\;\;
w = \frac{s'p}{dq},\;\;\;\lambda_0 = \frac{\mu'+\nu'}{d}.
\end{equation*}
Here $w$ is the maximum protein concentration in the active state. Taking inverse Laplace transform \cite{prudnikov1992integrals} gives rise to the steady-state protein distribution
\begin{equation}
p(x) = \frac{\Gamma(\beta)w^{1-\beta}}{\Gamma(\alpha)\Gamma(\beta-\alpha)\confluent(\alpha;\beta;w\lambda_0)}
x^{\alpha-1}(w-x)^{\beta-\alpha-1}e^{\lambda_0x},\;\;\;x<w,
\end{equation}
which is a beta-like distribution. In fact, this formula can also be obtained by solving the Kolmogorov backward equation \eqref{Kolmogorov} directly. However, this is much more difficult than our current method. In the switching ODE model, the protein concentration cannot exceed its maximum value $w$ and thus must be concentrated on $x<w$. Similarly, taking the limit of $K\rightarrow\infty$ in \eqref{activeprob} gives rise to the steady-state active probability of the gene \cite[Section 3]{supp}
\begin{equation*}
P_{\mathrm{active}} = \frac{\alpha}{\beta}
\frac{\confluent(\alpha+1;\beta+1;w\lambda_0)}{\confluent(\alpha;\beta;w\lambda_0)}.
\end{equation*}

We next focus on two special cases. When $b = \mu' = \nu' = 0$, the gene is always active and the protein concentration evolves as an ODE with fixed point $w$. In this case, the steady-state protein distribution reduces to the point mass at $w$, that is, $p(x) = \delta(x-w)$. When $\mu' = \nu' = 0$, the gene is unregulated. In this case, we have $\lambda_0 = 0$ and thus the protein concentration has the beta distribution
\begin{equation}
p(x) = \frac{\Gamma(\beta)w^{1-\beta}}{\Gamma(\alpha)\Gamma(\beta-\alpha)}x^{\alpha-1}(w-x)^{\beta-\alpha-1},\;\;\;x<w,
\end{equation}
where
\begin{equation*}
\alpha = \frac{a}{d},\;\;\;\beta = \frac{a+b}{d},\;\;\;w = \frac{s'p}{dq}.
\end{equation*}

To see the performance of the Kurtz limit, we numerically simulate both the discrete CME model using Gillespie's algorithm and the switching ODE model using the Euler-Maruyama scheme under two sets of biologically relevant parameters. Fig. \ref{kurtz}(b),(c) illustrate the steady-state distributions of the protein concentration for the two models. It can be seen that they agree with each other reasonably well when $K\gg 1$, but they fail as expected for smaller $K$. Both the two models can yield monomodal or bimodal steady-state protein distribution. Fig. \ref{kurtz}(b) corresponds to the monomodal case and Fig. \ref{kurtz}(c) corresponds to the bimodal case with the two modes peaking at $x = 0$ and $x = w$. Since the protein concentration in the Kurtz limit cannot exceed its maximal value $w$ while the discrete model does not have this constraint, the switching ODE model may deviate from the CME model significantly when the protein concentration is around $w$, even when $K$ is very large (Fig. \ref{kurtz}(c)).

\subsection{L\'{e}vy limit}
We next investigate the L\'{e}vy limit of the discrete CME model. To this end, we assume that the mean burst size scales with $K$ as $p/q = K/k$ and the feedback strengths scale with $1/K$ as $\mu = \mu'/K$ and $\nu = \nu'/K$, where we treat $s,d,k,a,b,\mu',\nu'$ as constants. Similarly, taking the limit of $K\rightarrow\infty$ in the CME \eqref{master} yields the following system of partial differential equations \cite[Section 4]{supp}:
\begin{equation*}\left\{
\begin{split}
\partial_tp_0(x) &= d\partial_x(xp_0(x))+(b+\nu'x)p_1(x)-(a+\mu'x)p_0(x),\\
\partial_tp_1(x) &= d\partial_x(xp_1(x))+s\int_0^x ke^{-k(x-y)}p_1(y)dy
+(a+\mu'x)p_0(x)-(b+s+\nu'x)p_1(x).
\end{split}\right.
\end{equation*}
From the viewpoint of stochastic processes, this is the Kolmogorov forward equation of the following switching SDE model driven by a compound Poisson process:
\begin{equation*}
\xymatrix{
\dot{x} = -dx+\dot\xi_t \ar@<0.2ex>@^{->}[d]^{b+\nu'x} & (\textrm{active gene}),\\
\dot{x} = -dx \ar@<0.2ex>@^{->}[u]^{a+\mu'x} & (\textrm{inactive gene}).}
\end{equation*}
Therefore, the L\'{e}vy limit of the discrete CME model is a switching SDE model. This is called the L\'{e}vy limit because the noise term $\xi_t$ is a compound Poisson process, a particular kind of L\'{e}vy process, with arrival rate $s$ and jump distribution $w(x) = ke^{-kx}$. This can be explained as follows. When the gene is active, the process of mRNA synthesis can be described by a Poisson process with arrival rate $s$ and each transcript can produce protein copies with the burst size having the exponential distribution $w(x)$, which can be viewed as the continuous limit of the geometric distribution. Then the process of protein synthesis should be described by the compound Poisson process $\xi_t$.

\begin{figure}[!htb]
\centerline{\includegraphics[width=1\textwidth]{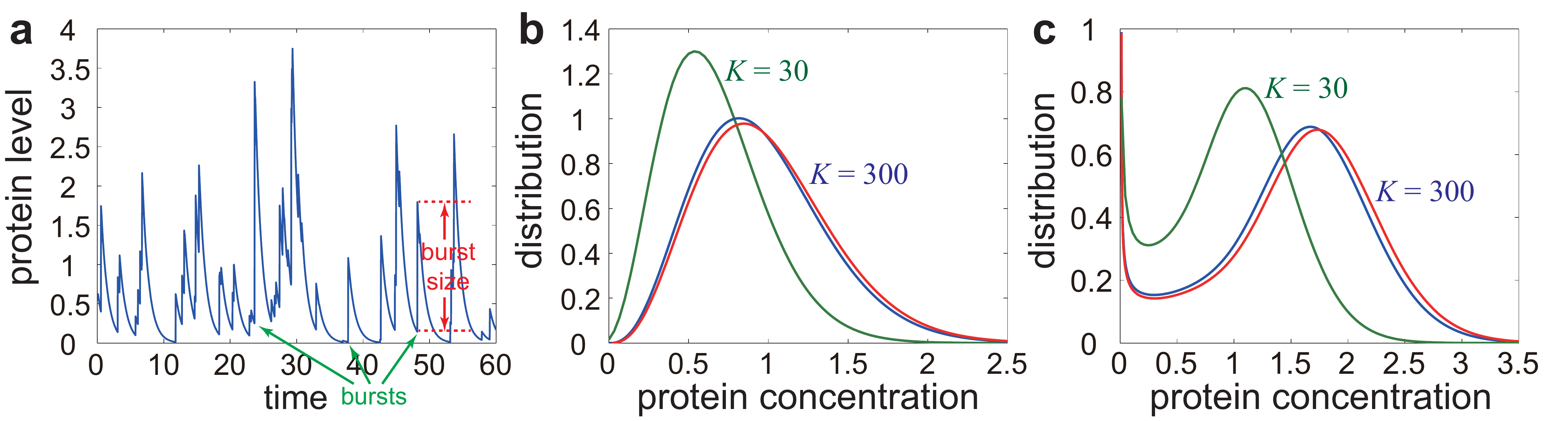}}
\caption{\textbf{Stochastic gene expression kinetics described by the L\'{e}vy limit.}
(a) The simulated trajectory of the switching SDE model. When the gene is active, the system evolves as an SDE with large fluctuations. The model parameters are chosen as $s = 2, d = 1, k = 2, p = 0.5, a = b = 1, \mu' = \nu' = 0.5$.
(b)-(c) Simulations of the steady-state distributions of the protein concentration for the switching SDE model (red) and the discrete CME model when $K = 30$ (green) and $K = 300$ (blue). (b) The case of monomodal protein distributions with the model parameters being chosen as $s = 20, d = 1, k = 10, a = b = 5, \mu' = \nu' = 1$. (c) The case of bimodal protein distributions with the model parameters being chosen as $s = 20, d = 1, k = 10, a = 0.6, b = 0.4, \mu' = 1, \nu' = 0$.}\label{levy}
\end{figure}

There is a crucial difference between the two macroscopic limits. Fig. \ref{levy}(a) illustrates the simulated time series of the protein concentration in the L\'{e}vy limit, where the model parameters are chosen so that the L\'{e}vy limit has the same mean field dynamics as the Kurtz limit depicted in Fig. \ref{kurtz}(a). Unlike the switching ODE model, the stochastic trajectories of the switching SDE model are discontinuous, where the jumps in each trajectory capture random translational bursts. The jump positions correspond to burst times and the jump heights correspond to burst sizes. Comparing Fig. \ref{kurtz}(a) with Fig. \ref{levy}(a), we clearly see that although the two macroscopic limits share the same mean field dynamics, the L\'{e}vy limit exhibits more drastic stochastic fluctuations. This is because the L\'{e}vy limit retains stochasticity coming from individual births and deaths of the mRNA and protein, while such stochasticity is averaged out in the Kurtz limit.

The biochemical implications of the two macroscopic limits can be understood as follows. Since the Kurtz limit is applicable when $s/d\gg 1$ and $p/q$ is finite, it corresponds to proteins with large burst frequencies. On the other hand, since the L\'{e}vy limit is applicable when $p/q\gg 1$ and $s/d$ is finite, it corresponds to proteins with large burst sizes. Recent single-cell experiments have shown that the burst sizes of many proteins are large, typically on the order of 100 for an \emph{E. coli} gene \cite{paulsson2005models}. This supports and justifies the L\'{e}vy limit that we have taken. In addition, we have seen that the stochastic trajectories of the L\'{e}vy limit are discontinuous with exponentially distributed jumps. This explains why the time-lapse measurements of expression levels of many proteins often display discontinuous trajectories with large jumps in single-cell time-lapse microscopy experiments \cite{suter2011mammalian}.

\subsection{Time-dependent solution of the Friedman-Cai-Xie model}
An important special case occurs when $b = \mu' = \nu' = 0$. In this case, the gene is always active and the evolution of the L\'{e}vy limit is governed by
\begin{equation}\label{friedman}
\partial_tp(x) = d\partial_x\big(xp(x)\big)+s\int_0^xw(x-y)p(y)dy-sp(x).
\end{equation}
This is exactly the classical Friedman-Cai-Xie (FCX) random bursting model proposed in \cite{friedman2006linking}, which is the Kolmogorov forward equation of the SDE
\begin{equation*}
\dot{x} = -dx+\dot\xi_t.
\end{equation*}
In fact, it has been shown that its steady-state solution is the gamma distribution \cite{friedman2006linking}. However, its time-dependent solution is still unknown up till now.

To obtain the time-dependent solution of the FCX model, let $\hat{p}(\lambda,t) = \int_0^\infty p(x,t)e^{-\lambda x}dx$ denote the Laplace transform of the time-dependent protein distribution. Then the FCX equation \eqref{friedman} can be transformed into the first-order linear partial differential equation
\begin{equation*}
\partial_t\hat{p} = -d\lambda\partial_\lambda\hat{p}-\frac{s\lambda\hat{p}}{\lambda+k}.
\end{equation*}
By using the method of characteristics, the solution of this partial differential equation is given by
\begin{equation*}
\hat{p}(\lambda,t) = \hat{p}(\lambda e^{-dt},0)\left(\frac{\lambda e^{-dt}+k}{\lambda+k}\right)^{s/d},
\end{equation*}
where $\hat{p}(\lambda,0) = \int_0^\infty p(x,0)e^{-\lambda x}dx$ is the Laplace transform of the initial protein distribution. Taking inverse Laplace transform \cite{prudnikov1992integrals}, we find that the time-dependent protein distribution is the convolution of two probability distributions $u$ and $v$:
\begin{equation}
p(x,t) = u*v(x,t) = \int_0^x u(x-y,t)v(y,t)dy,
\end{equation}
where $u(x,t) = e^{dt}p(e^{dt}x,0)$ and $v(x,t) = e^{-st}(w(x,t)+\delta(x))$ with $w(x,t)$ being defined as
\begin{equation*}
w(x,t) = \frac{sk}{d}(e^{dt}-1)e^{-ke^{dt}x}{}_1F_1(s/d+1;2;k(e^{dt}-1)x)I_{\{x\geq 0\}}.
\end{equation*}
Here $I_{\{x\geq 0\}}$ is the indicator function which takes the value of 1 when $x\geq 0$ and takes the value of 0 when $x<0$. In particular, if the initial protein concentration is $x_0$, then $u(x) = e^{dt}\delta(e^{dt}x-x_0)$ and thus the time-dependent protein distribution is given by
\begin{equation}
p(x,t) = e^{-st}[w(x-e^{-dt}x_0,t)+\delta(x-e^{-dt}x_0)],
\end{equation}
which is the sum of two parts:
\begin{equation*}
p_c(x,t) = e^{-st}w(x-e^{-dt}x_0,t),\;\;\;p_s(x,t) = e^{-st}\delta(x-e^{-dt}x_0).
\end{equation*}
This time-dependent solution has some interesting properties. First, it is clear that both the two parts vanish when $x<e^{-dt}x_0$. This can be explained as follows. We have shown that noise term $\xi_t$ captures random bursts of the protein. If the burst does not occur before time $t$, then the evolution of the protein concentration is governed by the deterministic dynamics $\dot{x} = -dx$, which undergoes an exponential decay with rate $d$. This implies that $e^{-dt}x_0$ is the minimum possible value of the protein concentration at time $t$. This explains why both the two parts vanishes when $x<e^{-dt}x_0$.

Second, both the two parts are discontinuous at $x=e^{-dt}x_0$. Specifically, the first part $p_c(x,t)$ has a jump at $x = e^{-dt}x_0$ with height $H = (sk/d)e^{-st}(e^{dt}-1)$ and the second part $p_s(x,t)$ is a constant multiple of a delta function, which has a spike at $x = e^{-dt}x_0$. The existence of a spike shows that at time $t$, there is a point mass $P = e^{-st}$ of the protein concentration at $x = e^{-dt}x_0$. This can be explained as follows. Since the L\'{e}vy limit is driven by a compound Poisson process with arrival rate $s$, the first burst time of the protein has an exponential distribution with rate $s$. Therefore, the probability that the burst does not occur before time $t$ is exactly $P = e^{-st}$. Provided that the burst does not occur before time $t$, the protein concentration undergoes an exponential decay with rate $d$. As a a result, there is a positive probability $P = e^{-st}$ for the protein concentration being exactly equal to $x = e^{-dt}x_0$ at time $t$.
\begin{figure}[!htb]
\centerline{\includegraphics[width=1\textwidth]{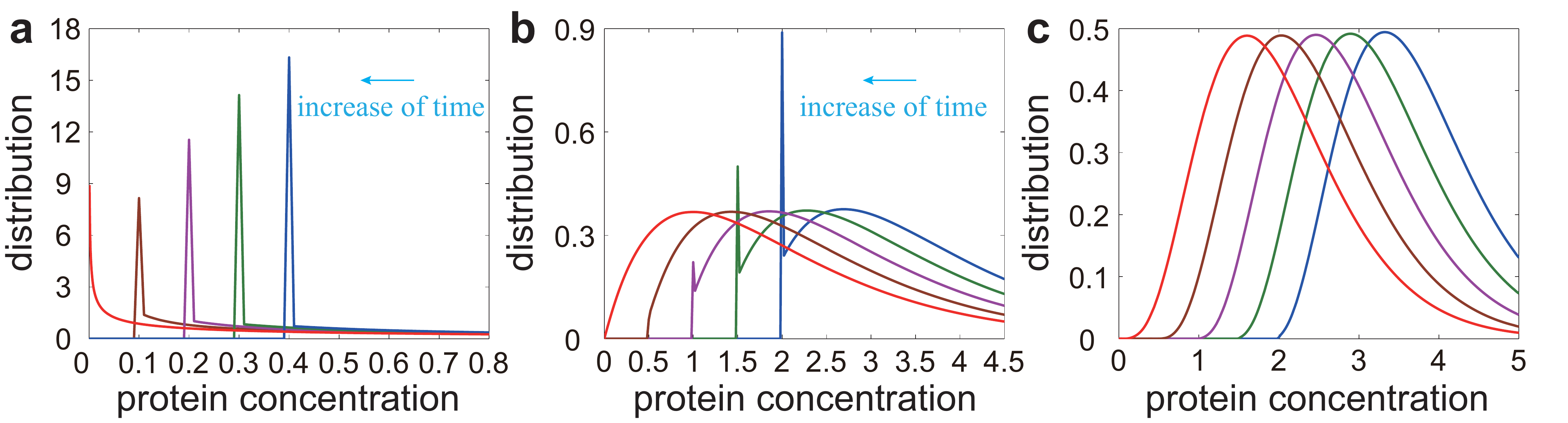}}
\caption{\textbf{Time-dependent solution of the FCX equation.}
(a)-(c) Simulations of the steady-state and time-dependent solutions of the FCX equation under four different choices of time points, where the red curve corresponds to the steady-state solution. In (a)-(c), the means of all steady-state protein distributions are kept to be $1$.
(a) The case of $s\leq d$, where the steady-state protein distribution is monotonically decreasing. The model parameters are chosen as $s = 0.5, k = s/2, d = 1, x_0 = 15$.
(b)-(c) The case of $s>d$, where the steady-state protein distribution is bell-shaped. In (b), the transcription rate $s$ is relatively small with the model parameters being chosen as $s = 2, k = s/2, d = 1, x_0 = 15$. In (c), the transcription rate $s$ is relatively large with the model parameters being chosen as $s = 5, k = s/2, d = 1, x_0 = 15$.}\label{FCX}
\end{figure}

To gain an intuitive picture of the time-dependent solution, we numerically simulate the SDE model using a method combining Gillespie's algorithm and the Euler-Maruyama scheme. In \cite{friedman2006linking}, the authors have shown that the steady-state protein distribution $p(x)$ of the FCX model may exhibit two types of behaviors: $p(x)$ peaks at zero when $s\leq d$ and has a nonzero peak when $s>d$. Fig. \ref{FCX}(a),(b) illustrate the steady-state and time-dependent protein distributions for the FCX model under four different time points in the cases of $s\leq d$ and $s>d$. In both cases, the time-dependent solutions display apparent spikes. As time increases, the position of the spike decreases exponentially with rate $d$ and the point mass of the spike decreases exponentially with rate $s$. These observations are fully consistent with our analytic results.

Finally, we notice that when $s\gg d$, both the jump height $H = (sk/d)e^{-st}(e^{dt}-1)$ of the first part $p_c(x,t)$ and the point mass $P = e^{-st}$ of the second part $p_s(x,t)$ are negligible whenever time $t$ is not very small. Fig. \ref{FCX}(c) depicts the steady-state and time-dependent protein distributions for the FCX model in the case of $s\gg d$, from which we can see that the time-dependent solutions are approximately continuous because of negligible jump heights and point masses.

\subsection{Steady-state protein distribution in the L\'{e}vy limit}
We next study the steady-state protein distribution for the switching SDE model of a coupled gene circuit. In analogy to the calculation in the switching ODE model, taking $z\rightarrow 1$ and $K\rightarrow\infty$ while keeping $\lambda = (1-z)K$ as a constant in the generating function \eqref{generating}, we obtain the Laplace transform of the steady-state protein distribution \cite[Section 4]{supp}
\begin{equation}
\hat{p}(\lambda) = \frac{\hyper(\alpha_1,\alpha_2;\beta;-w(\lambda-\lambda_0))}{\hyper(\alpha_1,\alpha_2;\beta;w\lambda_0)},
\end{equation}
where
\begin{gather*}
\alpha_1+\alpha_2 = \frac{a+b+s}{d},\;\;\;\alpha_1\alpha_2 = \frac{as}{d^2},\\
\beta = \frac{a+b}{d}+\frac{s\nu'}{d(\mu'+\nu'+dk)},\;\;\;
w = \frac{1}{\lambda_0+k},\;\;\;\lambda_0 = \frac{\mu'+\nu'}{d}.
\end{gather*}
Taking inverse Laplace transform \cite{prudnikov1992integrals} gives rise to the steady-state protein distribution
\begin{equation}
p(x) = \frac{\Gamma(\beta)w^{\frac{1-\alpha_1-\alpha_2}{2}}}
{\Gamma(\alpha_1)\Gamma(\alpha_2)\hyper(\alpha_1,\alpha_2;\beta;w\lambda_0)}
x^{\frac{\alpha_1+\alpha_2-3}{2}}e^{(\lambda_0-\frac{1}{2w})x}
W_{\frac{\alpha_1+\alpha_2+1}{2}-\beta,\frac{\alpha_1-\alpha_2}{2}}(x/w),
\end{equation}
where $W_{\alpha,\beta}(x)$ is the Whittaker function. Similarly, taking the limit of $K\rightarrow\infty$ in \eqref{activeprob} gives rise to the steady-state active probability of the gene \cite[Section 4]{supp}
\begin{equation*}
P_{\mathrm{active}} = \frac{ak}{\beta(\mu'+\nu'+dk)}
\frac{\hyper(\alpha_1+1,\alpha_2+1;\beta+1;w\lambda_0)}{\hyper(\alpha_1,\alpha_2;\beta;w\lambda_0)}.
\end{equation*}

We next focus on two special cases. When the gene is always active, that is, $b = \mu' = \nu' = 0$, the five parameters reduce to
\begin{equation*}
\alpha_1 = \beta = \frac{a}{d},\;\;\;\alpha_2 = \frac{s}{d},\;\;\;w = \frac{1}{k},\;\;\;\lambda_0 = 0,
\end{equation*}
and thus the Wittaker function reduces to \cite[Equation 13.18.2]{special}
\begin{equation*}
W_{\frac{\alpha_1+\alpha_2+1}{2}-\beta,\frac{\alpha_1-\alpha_2}{2}}(x/w)
= W_{\frac{\alpha_2-\alpha_1+1}{2},\frac{\alpha_1-\alpha_2}{2}}(kx)
= e^{-\frac{kx}{2}}(kx)^{\frac{\alpha_2-\alpha_1+1}{2}}.
\end{equation*}
In this case, the protein concentration has a gamma distribution, which is consistent with the result obtained by Friedman et al. \cite{friedman2006linking}:
\begin{equation}\label{gamma}
p(x) = \frac{k^{s/d}}{\Gamma(s/d)}x^{s/d-1}e^{-kx}.
\end{equation}
When $\mu' = \nu' = 0$, the gene is unregulated. In this case, we have $\lambda_0 = 0$ and thus the steady-state protein distribution can be simplified as
\begin{equation}
p(x) = \frac{k\Gamma(\beta)}{\Gamma(\alpha_1)\Gamma(\alpha_2)}
(kx)^{\frac{\alpha_1+\alpha_2-3}{2}}e^{-\frac{kx}{2}}
W_{\frac{\alpha_1+\alpha_2+1}{2}-\beta,\frac{\alpha_1-\alpha_2}{2}}(kx),
\end{equation}
where
\begin{equation*}
\alpha_1+\alpha_2 = \frac{a+b+s}{d},\;\;\;\alpha_1\alpha_2 = \frac{as}{d^2},\;\;\;
\beta = \frac{a+b}{d}.
\end{equation*}
Thus far, we have obtained the analytic expressions of the steady-state protein distributions for the discrete CME model and its two macroscopic limits. We summarize the corresponding distribution types in Table \ref{type}.
\begin{table}[!htb]\centering
\renewcommand\arraystretch{1.3}
\begin{tabular}{|c|c|c|c|} \hline
Conditions         & Discrete model      & Kurtz limit  & L\'{e}vy limit \\ \hline
General case       & Hypergeometric-type & Beta-like    & Wittaker-type \\ \hline
$\mu=\nu = 0$      & Hypergeometric-type & Beta         & Wittaker-type \\ \hline
$b =\mu=\nu = 0$   & Negative binomial   & Single-point & Gamma \\ \hline
\end{tabular}
\caption{The types of steady-state protein distributions for the discrete CME model and its two macroscopic limits.}\label{type}
\end{table}

To see the performance of the L\'{e}vy limit, we numerically simulate both the discrete CME model and the switching SDE model under two sets of biologically relevant parameters. Fig. \ref{levy}(b),(c) illustrate the steady-state distributions of the protein concentration for the two models. It can be seen that they coincide with each other perfectly when $K\gg 1$. However, the switching SDE model deviates from the discrete CME model when $K$ is relatively small. Both the two models can yield monomodal or bimodal steady-state protein distributions. Compared with the Kurtz limit, the L\'{e}vy limit behaves better in the bimodal case, especially for large values of the protein concentration.

\subsection{Significance of the macroscopic limits}
The two microscopic limits investigated above are important in several ways. First, they build a bridge between the discrete and continuous gene expression models. In recent years, switching ODE models \cite{ge2015stochastic, newby2015bistable, lin2016gene, bressloff2017stochastic, lin2018efficient, ge2018relatively} and switching SDE models \cite{friedman2006linking, mackey2013dynamic, jkedrak2016time, jia2017emergent} of stochastic gene expression kinetics have been extensively studied. However, the relationship between these continuous models the discrete CME model remains unclear. Our limit theory interlinks the discrete and continuous gene expression models by viewing the latter as the various microscopic limits of the former. This not only provides a rigorous theoretical foundation but also justifies the wide application for the switching ODE and SDE models, especially the classical FCX random bursting model.

In addition, the microscopic limits clarify the ranges of applicability of the switching ODE and SDE models. The former serves as a good approximation of the discrete CME model when the translational burst frequency is large, while the latter performs better when the translational burst size is large. For single-cell gene expression data with continuous measurements, it is more convenient to use continuous models rather than discrete models and our results provide insights into which continuous model should be selected.

Last but not least, the continuous models are often easier to handle than the discrete model. For instance, if we want to study bistable gene expression, the switching ODE model will be a good choice because the steady-state protein distribution for this model is a beta-like distribution, which can be represented by elementary functions. However, the steady-state protein distributions for the CME model and the switching SDE model contain hypergeometric and Wittaker functions, whose monotonicity and shapes are difficult to analyze. Recently, the switching ODE model has also been applied to provide an analytic theory of stochastic biochemical oscillations and the switching SDE model has been applied to analyze the influence of random bursts on stochastic oscillations \cite{jia2019analytic}. There are also some other applications of the microscopic limits. Since all stochasticity of the switching ODE model comes from promoter switching, the protein noise of this model is used by some authors to define the promoter switching noise \cite{liu2016decomposition}.

\section{Discussion}
In this work, we present a detailed analysis of single-cell stochastic gene expression kinetics in a minimal coupled gene circuit with positive-plus-negative feedback. Our theory builds a bridge between various discrete and continuous gene expression models proposed in the previous literature by viewing the latter as macroscopic limits of the former. Following \cite{jia2017emergent}, we focus on two types of macroscopic limits: the Kurtz limit applies in the regime of large burst frequencies and the L\'{e}vy limit applies in the regime of large burst sizes. The former turns out to be a switching ODE whose all stochasticity comes from promoter switching, while the latter turns out to be a switching SDE driven by L\'{e}vy noise which captures random translational bursts.

In the presence of coupled positive-plus-negative feedback loops, we assume that the promoter switching rates depend linearly on the protein copy number $n$ as $a_n = a+\mu n$ and $b_n = b+\nu n$. In fact, this assumption is equivalent to the following four chemical reactions:
\begin{gather*}
\textrm{inactive gene} \xlongrightarrow{a} \textrm{active gene},\\
\textrm{inative gene}+\textrm{protein} \xlongrightarrow{\mu} \textrm{active gene},\\
\textrm{active gene} \xlongrightarrow{b} \textrm{inactive gene},\\
\textrm{active gene}+\textrm{protein} \xlongrightarrow{\nu} \textrm{inactive gene}.
\end{gather*}
It is worth noting that if the second or fourth reaction occurs, the protein copy number should decrease by 1. However, in the CME model depicted in Fig. \ref{model}(b), we implicitly assume that when a protein copy binds to a gene, there is no change in the protein copy number. This is a small approximation made in the present paper and many previous papers, as pointed out by \cite{grima2012steady}. With this approximation, we calculate the steady-state protein distributions for the discrete CME model and its two macroscopic limits by using the methods of generating functions and Laplace transforms. These analytic distributions cover and extend most analytic results obtained in previous studies. When the gene is always active, the switching SDE model reduces to the classical FCX random bursting model \cite{friedman2006linking} and the present work also provides its time-dependent protein distribution.

Our analytic results are then applied to investigate the structure of gene expression noise in coupled gene circuits. The idea of decomposing noise in terms of different biophysical origins was first proposed by Paulsson \cite{paulsson2004summing}. Different types of noise could provide living organisms alternative mechanisms to improve fitness and control noise in fluctuating environments. If a gene is unregulated, a three-term noise decomposition into the protein birth-death noise, mRNA noise, and promoter switching noise has been proposed \cite{pedraza2008effects, shahrezaei2008analytical}. In the presence a positive or negative feedback loop, another three-term noise decomposition into the protein birth-death noise, mRNA noise, and feedback noise has been proposed in the regime of fast promoter switching \cite{jia2017stochastic}. In the regime of slow promoter switching, it is difficult to decompose gene expression noise due to the strong interaction between promoter switching and feedback regulation. In a recent work of Liu et al. \cite{liu2016decomposition}, the authors ignored the transcription dynamics and proposed an alternative noise decomposition into the protein birth-death noise, promoter switching noise, and correlation noise in the regime of slow promoter switching. However, their protein birth-death noise is not compatible with the decompositions in previous papers \cite{shahrezaei2008analytical}.

In the present work, we propose a complete five-term noise decomposition for coupled gene circuits under a wide range of biologically relevant parameters, which provides novel insights into how and to what extent coupled feedback loops can enhance or suppress molecular fluctuations. In addition to the protein birth-death noise, mRNA noise, and promoter switching noise, our decomposition gives the quantitative characterization of the contributions caused by positive and negative feedback loops. In fact, previous results have shown that positive feedback amplifies noise and negative feedback reduces noise in the regime of fast promoter switching, regardless of the feedback strengths \cite{jia2017stochastic}. Our results show that this conclusion is also valid in the regime of slow promoter switching when the feedback strengths are small. This result is expected to be also true when the feedback strengths are large if the promoter switching noise can be defined reasonably, but a rigorous mathematical theory is still lacking.

According to our analysis, the positive and negative feedback effects in a coupled gene circuit in general cannot be cancelled out. We discover that a coupled gene circuit undergoes a triphasic stochastic bifurcation as the ratio of the positive and negative feedback strengths increases. When the ratio is very large (small), the coupled gene circuit amplifies (diminishes) both the gene expression mean and gene expression noise and behaves like a positive (negative) feedback circuit. However, when the ratio is neither too small nor too large, a coupled gene circuit behaves neither like a positive feedback nor like a negative feedback circuit. Our model predicts that coupled positive-plus-negative feedback amplifies gene expression mean but diminish gene expression noise over a wide range of feedback strengths when promoter switching is relatively slow. This reveals a crucial difference between coupled feedback loops and a single feedback loop. Compared with a negative feedback circuit which stabilizes gene expression around a relatively low level and a positive feedback circuit which does not stabilize gene expression, a coupled gene circuit could stabilize gene expression around a relatively high level.

From the theoretical point of view, a future challenge is to extend our current analytic results to the steady-state joint distribution $p(n_1,n_2)$ of a protein pair. From the practical point of view, another future challenge is to link our stochastic kinetic approach to statistical or machine learning approach in order to obtain a better and more robust statistical inference of the model parameters from massive single-cell experimental data.

\section*{Acknowledgements}
The authors thank Professor Hong Qian and Professor Min Chen for stimulating discussions. The authors are also grateful to the anonymous referees for their valuable comments and suggestions which helped us greatly in improving the quality of this paper. C. Jia, L.Y. Wang, and G. Yin were supported in part by the Army Research Office under W911NF-19-1-0176. M.Q. Zhang was supported by NIH grants MH102616 and MH109665 and also by NSFC 31671384 and 91329000.

\setlength{\bibsep}{5pt}
\small\bibliographystyle{nature}

\end{document}